\documentclass[twocolumn,superscriptaddress]{revtex4-2}
\usepackage[colorlinks=true, citecolor=blue, urlcolor=blue, linkcolor=red]{hyperref}
\usepackage{amsmath,amssymb,tikz,scalerel}
\hypersetup{
        colorlinks=true,
        linkcolor=red,
        citecolor=blue,
        urlcolor=blue}
\usetikzlibrary{svg.path}
\definecolor{orcidlogocol}{HTML}{A6CE39}
\tikzset{
	orcidlogo/.pic={
		\fill[orcidlogocol] svg{M256,128c0,70.7-57.3,128-128,128C57.3,256,0,198.7,0,128C0,57.3,57.3,0,128,0C198.7,0,256,57.3,256,128z};
		\fill[white] svg{M86.3,186.2H70.9V79.1h15.4v48.4V186.2z}
		svg{M108.9,79.1h41.6c39.6,0,57,28.3,57,53.6c0,27.5-21.5,53.6-56.8,53.6h-41.8V79.1z M124.3,172.4h24.5c34.9,0,42.9-26.5,42.9-39.7c0-21.5-13.7-39.7-43.7-39.7h-23.7V172.4z}
		svg{M88.7,56.8c0,5.5-4.5,10.1-10.1,10.1c-5.6,0-10.1-4.6-10.1-10.1c0-5.6,4.5-10.1,10.1-10.1C84.2,46.7,88.7,51.3,88.7,56.8z};}}
\newcommand\orcid[1]{\href{https://orcid.org/#1}{\mbox{\scalerel*{\begin{tikzpicture}[yscale=-1,transform shape]\pic{orcidlogo};\end{tikzpicture}}{|}}}}

\begin{document}
\title{Quantum metrology in the noisy intermediate-scale quantum era}
\author{Lin Jiao}
\author{Wei Wu\orcid{0000-0002-7984-1501}}
\affiliation{Key Laboratory of Quantum Theory and Applications of MoE, Lanzhou Center for Theoretical Physics, and Key Laboratory of Theoretical Physics of Gansu Province, Lanzhou University, Lanzhou 730000, China}
\author{Si-Yuan Bai\orcid{0000-0002-4768-6260}}
\affiliation{Key Laboratory of Quantum Theory and Applications of MoE, Lanzhou Center for Theoretical Physics, and Key Laboratory of Theoretical Physics of Gansu Province, Lanzhou University, Lanzhou 730000, China}
\author{Jun-Hong An\orcid{0000-0002-3475-0729}}
\email{anjhong@lzu.edu.cn}
\affiliation{Key Laboratory of Quantum Theory and Applications of MoE, Lanzhou Center for Theoretical Physics, and Key Laboratory of Theoretical Physics of Gansu Province, Lanzhou University, Lanzhou 730000, China}
\begin{abstract}
Quantum metrology pursues the physical realization of higher-precision measurements to physical quantities than the classically achievable limit by exploiting quantum features, such as entanglement and squeezing, as resources. It has potential applications in developing next-generation frequency standards, magnetometers, radar, and navigation. However, the ubiquitous decoherence in the quantum world degrades the quantum resources and forces the precision back to or even worse than the classical limit, which is called the no-go theorem of noisy quantum metrology and greatly hinders its applications. Therefore, how to realize the promised performance of quantum metrology in realistic noisy situations attracts much attention in recent years. We will review the principle, categories, and applications of quantum metrology. Special attention will be paid to different quantum resources that can bring quantum superiority in enhancing sensitivity. Then, we will introduce the no-go theorem of noisy quantum metrology and its active control under different kinds of noise-induced decoherence situations.

\end{abstract}

\maketitle

\section{Introduction}

Quantum metrology aims at achieving highly precise measurement of a physical quantity of interest with the help of certain quantum resources as well as the principles of quantum mechanics~\cite{doi:10.1126/science.1104149,PhysRevLett.96.010401,RevModPhys.90.035005}. These resources have no classical counterparts and result in quantum superiority over traditional metrology schemes. The sensitivity of classical metrology schemes is commonly constrained by the so-called shot-noise limit (SNL) $\delta\theta\propto N^{-1/2}$, where $\delta\theta$ is the root-mean-square error of the quantity $\theta$ and $N$ is the number of repeated experiments. The above SNL is guaranteed by the central limit theorem in physical statistics. By increasing the number of repeated experiments, the metrological error can be reduced. In this sense, $N$ is viewed as the number of resources in classical schemes. Generalizing to the quantum case, the concept of the number of resources is greatly extended. It has been demonstrated that if quantum resources, such as entanglement~\cite{doi:10.1116/1.5120348,PhysRevLett.125.100402,doi:10.1063/5.0050235,PhysRevA.95.023608} or squeezing~\cite{PhysRevA.46.R6797,PhysRevA.47.5138,PhysRevLett.127.083602}, are used, the metrological sensitivity can surpass the classical SNL~\cite{Giovannetti1330,PhysRevA.46.R6797,PhysRevLett.69.3598,PhysRevLett.71.1355,PhysRevLett.96.010401,RevModPhys.89.035002,RevModPhys.90.035005}. Thus, the numbers of the entangled particles, the squeezing parameter, and the sizes of the many-body systems experiencing a quantum phase transition, are generally regarded as the number of quantum resources. It has been found that the metrology precision with a $N$-body entangled state reaches the Heisenberg limit (HL) scaling $N^{-1}$, which beats the SNL $N^{-1/2}$ times. Such an enhancement
inspires many fantastic applications in gravitational-wave detection~\cite{Schnabel2010,PhysRevLett.110.181101,PhysRevD.94.124043}, quantum radar~\cite{PhysRevLett.114.080503,PhysRevLett.124.200503,PhysRevResearch.2.023414}, atom clocks~\cite{PhysRevLett.117.143004,Hosten505,PhysRevLett.125.210503,PhysRevX.11.041045}, magnetometers~\cite{Jonathan2009,PhysRevLett.113.103004,RevModPhys.89.035002,PhysRevLett.125.020501}, gravimeter~\cite{PhysRevLett.117.138501,PhysRevLett.118.183602},  navigation~\cite{PhysRevLett.111.230503,doi:10.1116/1.5120348}, and biological monitoring~\cite{TAYLOR20161,doi:10.1063/1.4724105,PhysRevX.4.011017}.

Quantum metrology is on the way to enhancing its practical capacity to outperform the state-of-the-art classical metrology counterparts. The gravitational-wave detection has achieved an enhanced sensitivity over the SNL by using the squeezed light \cite{PhysRevLett.123.231107,PhysRevLett.123.231108}. However, in the noisy intermediate-scale quantum (NISQ) era, the realization of quantum metrology is generally challenged by decoherence caused by different kinds of noises. In any practical quantum metrology scheme, the quantum probe unavoidably interacts with its surrounding noises and becomes an open system~\cite{PhysRevA.81.033819,PhysRevA.90.033846,PhysRevLett.102.040403,PhysRevA.80.013825,PhysRevA.95.053837,PhysRevA.78.063828,Gilbert:08,PhysRevA.75.053805}. Its evolution impairs the quantum coherence, which is called decoherence. Quantum resources are very fragile and can be easily destroyed by decoherence. Decoherence results in the deterioration of the performance of quantum metrology. It was found that the metrology error generally returns to the SNL at an optimal transient encoding time and becomes divergent in the long-encoding-time regime under the influence of decoherence ~\cite{Haase_2018,PhysRevA.97.012125,Tamascelli_2020,PhysRevLett.98.160401,PhysRevApplied.5.014007,PhysRevA.99.033807,PhysRevA.102.012223,PhysRevResearch.2.033389}. Such a phenomenon is called the no-go theorem of noisy quantum metrology \cite{PhysRevLett.116.120801} and is the main obstacle to achieve a high-precision quantum metrology in practice. However, a clear imperfection leading to this no-go theorem is that
it is based on the Born-Markovian approximation to describe the decoherence. Thus, the determination of whether this no-go theorem is ostensible or fundamental and can be overcome is highly desirable from both theoretical and experimental perspectives.

The potential benefit of an advanced technology innovation offered by quantum metrology makes overcoming the challenge set by decoherence an extremely worthwhile goal. To minimize the unwanted effects of decoherence on quantum metrology, various strategies have been proposed. Many efforts, e.g., adaptive \cite{PhysRevLett.111.090801,PhysRevX.7.041009} and nondemolition \cite{PhysRevLett.125.200505} measurements, correlated decoherence \cite{Jeske_2014}, purification \cite{PhysRevLett.129.250503}, error correction \cite{PhysRevLett.112.150802,PhysRevLett.112.080801,PhysRevLett.112.080801,PhysRevLett.116.230502,Lu2015,Reiter2017,Zhou2018,PhysRevLett.128.140503}, and dynamical control \cite{PhysRevA.87.032102,PhysRevA.94.052322,Sekatski_2016}, have been proposed to restore the HL. A mechanism of quantum reservoir engineering to solve the divergence problem was proposed in \cite{Wang_2017,PhysRevLett.123.040402,PhysRevA.104.042609,Jiao:23}. A Floquet engineering to retrieve the HL and overcome the precision divergence with time simultaneously was reported in \cite{PhysRevLett.131.050801}. With advancements in different strategies for mitigating decoherence effects, quantum metrology is on the brink of transitioning from the laboratory to practical use. In this paper, we give a brief review of the decoherence control schemes to realize a high-precision metrological performance.

This paper is organized as follows. In Sec.~\ref{chap:Q.estim}, we review the basic concepts as well as the general formalism of quantum parameter
estimation. The Fisher information as well as two kinds of Cram\'{e}r-Rao bound are introduced. In Sec.~\ref{chap:resources}, we review the categories and their respective physical principles of the ideal quantum-metrology schemes. The widely used quantum-metrology schemes based on Ramsey spectroscopy, Mach-Zehnder interferometer, and Sagnac interferometer are introduced. In Sec.~\ref{otqsr}, we review the quantum resources other than the entanglement and the squeezing, including spin squeezing, quantum criticality, and quantum chaos, which are useful in enhancing the sensitivity of quantum metrology. The decoherence effects on quantum metrology are reviewed in Sec. \ref{effnqm}. In Sec.~\ref{control}, the widely used control schemes to suppress decoherence in noisy quantum metrology are discussed. The conclusions and outlook of this paper are drawn in Sec.~\ref{conclusion}.

\section{Quantum parameter estimation}\label{chap:Q.estim}


Any quantum metrology scheme generally includes three steps. To measure an unknown quantity $\theta$ of a physical system, we first prepare a quantum probe in specific state $\varrho_\text{in}$ containing a certain quantum resource. Then, we couple the probe to the system to encode $\theta$ into the probe state via a dynamical mapping as $\varrho_{\theta}=\hat{\Lambda}_{\theta}(\varrho_{\text{in}})$, where the mapping operator $\hat{\Lambda}_{\theta}$ may be either unitary or nonunitary. Finally, we measure certain observable $\hat{O}\equiv\sum_{j}o_j|o_j\rangle\langle o_j|$ of the probe in $\varrho_{\theta}$ and infer the value of $\theta$ from the result. The measurement yields an outcome $o_j$ with a probability $p(j|\theta)=\langle o_j|\varrho_\theta|o_j\rangle$. The metrology sensitivity of $\theta$ is constrained by the famous Cram\'{e}r-Rao bound~\cite{Stoica17564}
\begin{equation}\label{eq:eq2-1}
\delta^{2}\theta\geq (\upsilon F_{\theta})^{-1},
\end{equation}
where $\delta\theta$ is the root-mean-square error of $\theta$, $\upsilon$ is the number of repeated experiments, and $F_{\theta}$ is the classical Fisher information (CFI) corresponding to the selected measurement observable~\cite{PhysRevA.62.012107,Facchi2010,Liu_2020}. The CFI can be evaluated from the probability distribution $p(j|\theta)$ as
\begin{equation}\label{eq:eq2-2}
F_{\theta}=\sum_{j}p(j|\theta)\Big{[}\frac{\partial}{\partial\theta}\ln p(j|\theta)\Big{]}^{2}.
\end{equation}
The above projective measurement can be generalized to any positive-operator valued measure $\{\hat{\Pi}_{j}^\dag\hat{\Pi}_{j}\}$, which satisfies $\sum_{j}\hat{\Pi}_{j}^{\dagger}\hat{\Pi}_{j}=\mathbf{I}$. The corresponding probability distribution reads $p(j|\theta)=\mathrm{Tr}(\hat{\Pi}_{j}\varrho_{\theta}\hat{\Pi}_{j}^{\dagger})$.

Equation~(\ref{eq:eq2-2}) reveals that the CFI strongly relies on the choice of measured observable. Optimizing all the possible measurement observables, the ultimate metrology sensitivity of $\theta$ is constrained by the quantum Cram\'{e}r-Rao bound
\begin{equation}\label{eq:eq2-3}
\delta^{2}\theta\geq (\upsilon \mathcal{F}_{\theta})^{-1},
\end{equation}
where $\mathcal{F}_{\theta}\equiv \mathrm{Tr}(\hat{\varsigma}^{2}\varrho_{\theta})$ is the so-called quantum Fisher information (QFI)~\cite{Facchi2010,Liu_2020}. $\hat{\varsigma}$ is the symmetric logarithmic derivative and determined by $\partial_{\theta}\varrho_{\theta}=\frac{1}{2}(\hat{\varsigma}\varrho_{\theta}+\varrho_{\theta}\hat{\varsigma})$.
The QFI describes the maximal information on $\theta$ extractable from $\varrho_{\theta}$ optimizing all possible measurement observables, while the CFI denotes the maximal information extractable from the selected measurement observable. Thus, one can immediately conclude that $\mathcal{F}_{\theta}\geq F_{\theta}$. If the selected measurement scheme is the optimal one, then $F_{\theta}$ would saturate to $\mathcal{F}_{\theta}$. Unfortunately, there is no general way to find the optimal measurement observable. In this sense, designing the physically optimal measurement scheme that can saturate the best attainable precision bounded by the QFI is of importance in the study of quantum metrology. Some progress has been achieved to indirectly extract the QFI by measuring the energy fluctuation~\cite{PhysRevA.107.012414}.

In some studies, instead of CFI, the metrological sensitivity is alternatively quantified by using the so-called error propagation formula
\begin{equation}
\delta^{2}\theta=\frac{\Delta^{2}O}{|\partial\langle\hat{O}\rangle/\partial\theta|^{2}},\label{epg}
\end{equation}
where $\langle\hat{O}\rangle=\text{Tr}(\hat{O}\varrho_{\theta})$ and $\Delta^{2}O=\langle\hat{O}^{2}\rangle-\langle\hat{O}\rangle^{2}$ are the expectation value and the fluctuation of the chosen operator $\hat{O}$ in the state $\varrho_{\theta}$, respectively. Compared with that of CFI, the error propagation formula provides a convenient way to characterize the performance of a given quantum metrology scheme. Another quantity to reflect the sensitivity under certain measurement scheme is the signal-to-noise ratio (SNR) \cite{PhysRevLett.128.033601}
\begin{equation}
\text{SNR}_{\theta}=\frac{\langle\hat{O}\rangle}{\Delta O/\sqrt{\upsilon}}.\label{snnd}
\end{equation}
Both Eqs. \eqref{epg} and \eqref{snnd} can faithfully reflect the metrological sensitivity of a given scheme, but they do not correspond to the highest precision. By optimizing the chosen observable $\hat{O}$, they approach the ultimate precision determined by the QFI.

When the probe is a discrete-variable system, its density matrix $\varrho_{\theta}$ is conveniently described in the Bloch representation as
\begin{equation}
\varrho_{\theta}=d^{-1}\big{[}\mathbf{1}_{d}+\sqrt{{d(d-1)}/{2}}\pmb{r}\cdot\pmb{\zeta}\big{]},
\end{equation}
with $d$ being the dimension of the probe, $\pmb{r}=\text{Tr}(\pmb{\zeta}\varrho_\theta)$ being the Bloch vector, and $\pmb{\zeta}$ being the $(d^{2}-1)$-dimensional vector of the generators of the group SU$(d)$. The QFI is calculated via~\cite{Liu_2020}
\begin{equation}\label{eq:eq2-41}
\mathcal{F}_{\theta}=(\partial_{\theta}\pmb{r})^{\text{T}}\cdot\big[\frac{d}{2(d-1)}\pmb{G}-\pmb{r}\cdot\pmb{r}^{\text{T}}\big]^{-1}\cdot\partial_{\theta}\pmb{r},\end{equation}
where $\pmb{G}$ is a real symmetric matrix with
\begin{equation}
\pmb{G}_{ij}=\frac{1}{2}\text{Tr}\big{(}\varrho_{\theta}\{\zeta_{i},\zeta_{j}\}\big{)}.
\end{equation}
The most used scenario is the single-qubit case with $d=2$. Under this circumstance, the QFI reduces to
\begin{equation}\label{eq:eq2-4}
\mathcal{F}_{\theta}=|\partial_{\theta}\pmb{r}|^{2}+\frac{(\pmb{r}\cdot\partial_{\theta}\pmb{r})^{2}}{1-|\pmb{r}|^{2}}.
\end{equation}
for a mixed state, and $\mathcal{F}_{\theta}=|\partial_{\theta}\pmb{r}|^{2}$ for a pure state. Here, $\pmb{\zeta}=(\hat{\sigma}_{x},\hat{\sigma}_{y},\hat{\sigma}_{z})^{\text{T}}$ reduces to the Pauli matrices.

If the probe is a Gaussian bosonic system with a set of annihilation and creation operators $\pmb{A}=\{\hat{a}_{1},..,\hat{a}_{N},\hat{a}_{1}^{\dagger},...\hat{a}_{N}^{\dagger}\}$, its quantum state $\varrho_{\theta}$ can be fully characterized by the first-order moments of the displacement vector $\pmb{d}$ and the second-order moments of the covariant matrix $\pmb{\sigma}$ ~\cite{PhysRevA.103.L010601}. The elements of ${\pmb{d}}$ and ${\pmb\sigma}$ are, respectively, defined by $\pmb{d}_{i} =\text{Tr}(\rho_\theta \pmb{A}_{i})$ and $\pmb{\sigma}_{ij} =\text{Tr}[\rho_\theta \{\Delta \pmb{A}_{i},\Delta \pmb{A}_{j}\}]$
with $\Delta \pmb{A}_{i}=\pmb{A}_{i}-\pmb{d}_{i}$. With expressions of $\pmb{d}$ and $\pmb{\sigma}$ at hand, the QFI with respect to the mixed Gaussian state $\varrho_{\theta}$ is calculated as~\cite{Gao2014,Safranek_2015,Safranek_2019}
\begin{equation}\label{eq:eq2}
\mathcal{F}_{\theta}=\frac{1}{2}[\text{vec}(\partial_\theta {\pmb \sigma})]^\dag\pmb{M}^{-1}\text{vec}(\partial_\theta {\pmb \sigma})+2(\partial_{\theta}\pmb{d})^{\dagger}\pmb{\sigma}^{-1}\partial_{\theta}\pmb{d},
\end{equation}
where $\text{vec}(\cdot)$ denotes the vectorization of a given matrix, and $\pmb{M}={\pmb\sigma}^*\otimes{\pmb\sigma}-\pmb{\varpi}\otimes \pmb{\varpi}$ with $[\pmb{A}_{i},\pmb{A}_{j}]=i\pmb{\varpi}_{ij}$. In the pure state case, the QFI reduces to
\begin{equation}
\mathcal{F}_{\theta}=\frac{1}{4}\text{Tr}[(\pmb{\sigma}^{-1}\partial_\theta {\pmb \sigma})^{2}]+2(\partial_{\theta}\pmb{d})^{\dagger}\pmb{\sigma}^{-1}\partial_{\theta}\pmb{d}.
\end{equation}

\section{Ideal quantum metrology}\label{chap:resources}

Depending on the feature of the Hilbert space of the probe, quantum metrology schemes are classified into two categories: discrete- and continuous-variable schemes. The discrete-variable quantum metrology scheme is generally based on Ramsey spectroscopy and the continuous-variable one is based on the Mach-Zehnder and Sagnac interferometers.

\subsection{Ramsey spectroscopy}\label{rmsspc}

\begin{figure}[tbp]
	\centering
	\includegraphics[width=1\columnwidth]{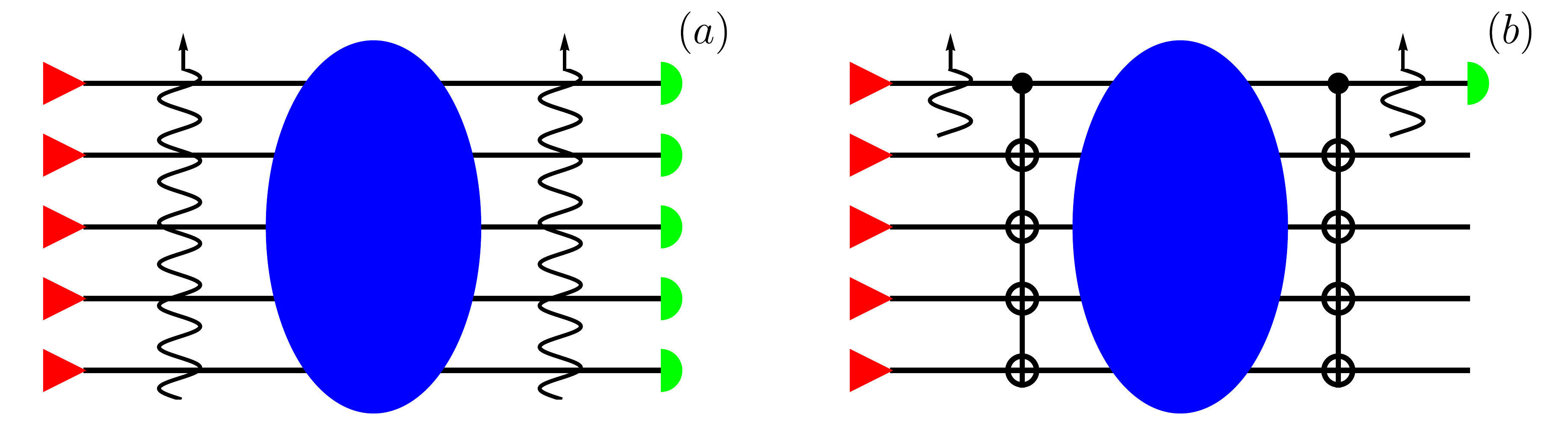}
	\caption{ Scheme of Ramsey-spectroscopy-based quantum metrology. (a) Conventional Ramsey spectroscopy uses a product state as the input state. (b) Ramsey spectroscopy uses a GHZ-type entangled state as the input state. }\label{RMSI}
\end{figure}

In a conventional Ramsey spectroscopy to measure the atomic frequency $\omega_0$, see Fig. \ref{RMSI}(a), one chooses $N$ atoms themselves as the probe and prepares their state in $|\psi_\text{in}\rangle=|g\rangle^{\otimes N}$, where $|g\rangle$ is the atomic ground state. First, a $\pi/2$ microwave pulse in frequency $\omega_L$ and time duration $t=\pi/(2|\Delta|)$, with $\Delta=\omega_0-\omega_L$, is applied on each atom. It converts $|\psi_\text{in}\rangle$ into $|\psi_1\rangle=\big({|g\rangle+|e\rangle\over\sqrt{2}}\big)^{\otimes N}$, where $|e\rangle$ is the excited state. Second, the microwave is switched off and the atoms experience a free evolution governed by the Hamiltonian $\hat{H}_{0}=\Delta\sum_j\hat{\sigma}_j^\dag\hat{\sigma}_j$, with $\hat{\sigma}_j=|g_j\rangle\langle e_j|$. It encodes $\omega_0$ into the state \begin{equation}
    |\psi(t)\rangle=e^{-i\hat{H}_{0}t}|\psi_1\rangle=[(|g\rangle+e^{-i\Delta t}|e\rangle)/\sqrt{2}]^{\otimes N}.
\end{equation} Here, $\omega_L$ contained in $\Delta$ of $\hat{H}_0$ is just for the purpose of the consistency of the used rotating frame with the one in the first step. Third, a $\pi/2$ microwave pulse is switched on again, which converts $|\psi(t)\rangle$ into
\begin{equation}
|\psi_\text{out}\rangle=\big[\cos(\Delta t/2)|g\rangle+i\sin(\Delta t/ 2)|e\rangle\big]^{\otimes N}.\label{resprdt}
\end{equation}
Finally, the excited-state population operator $\hat{O}=\hat{\sigma}_j^\dag\hat{\sigma}_j$ of each atom is measured. The obtained results 1 is in a probability $P_1=\sin^2{\Delta t\over 2}$ and 0 in a probability $P_0=\cos^2{\Delta t\over 2}$. The expectation value $\langle \hat{O}\rangle=\sin^2{\Delta t\over 2}$, the variance $\Delta O=|\sin\Delta t|/2$, and the CFI $F_{\omega_0}=t^2$ are readily calculated. Repeating the experiment within a time duration $T$, we acquire $\upsilon=TN/t$ measurement results. Then, according to the central limit theorem, we have the uncertainty of $\hat{O}$ as
\begin{equation}
\delta O={\Delta O\over \sqrt{TN/t}}={|\sin\Delta t|\over 2\sqrt{TN/t}}.
\end{equation}
The sensitivity of $\omega_0$ is evaluated via the error propagation formula as
\begin{equation}
\delta\omega_0={\delta O\over |\partial_{\omega_0}\langle\hat{O}\rangle|}=(NTt)^{-1/2},\label{rmssnl}
\end{equation}which equals to the sensitivity evaluated from the Cram\'{e}r-Rao bound \eqref{eq:eq2-1}.
Since no quantum resource is used in the input state, the scaling relation of $\delta\omega_0$ with the atom number $N$ is just the SNL. It can be evaluated that the QFI of Eq. \eqref{resprdt} is $\mathcal{F}_{\omega_0}=t^2$. Thus, we obtain
\begin{equation}
\delta\omega_0=(\upsilon\mathcal{F}_{\omega_0})^{-1/2}=(NTt)^{-1/2}.
\end{equation}
Therefore, the measurement scheme used in Eq. \eqref{rmssnl} saturates the quantum Cram\'{e}r-Rao bound and is optimal.

Entanglement is one of the most subtle and intriguing phenomena in the microscopic world. Its usefulness has been demonstrated in various applications such as quantum teleportation~\cite{PhysRevLett.84.4236,Bouwmeester1997}, quantum cryptography~\cite{RevModPhys.81.865,PhysRevLett.84.4729,Pirandola:20}, and quantum dense coding~\cite{RevModPhys.81.865,doi:10.1080/09500349514551091,doi.org/10.1002/qute.201900011}. It occurs when a group of particles are generated or share spatial proximity in a way such that their state cannot be described independently by the states of the constituent particles. Measurements of physical quantities such as position, momentum, spin, and polarization performed on entangled particles can be found to be perfectly correlated. The exploitation of entanglement in quantum technologies has great potential to outperform the schemes based on classical physics. This is also true for quantum metrology.

First, the $N$ atoms are prepared in the ground state $|\psi_\text{in}\rangle=|g\rangle^{\otimes N}$. A $\pi/2$ microwave pulse on the first atom is applied, see Fig. \ref{RMSI}(b), and converts the state into
$|\psi_1\rangle=\frac{|g\rangle+|e\rangle}{\sqrt{2}}\otimes |g\rangle^{\otimes (N-1)}$. A CNOT gate, with the first atom acting as the control bit and other $N-1$ atoms acting as the target bits, is applied to the atoms. $|\psi_1\rangle$ becomes a Greenberger–Horne–Zeilinger (GHZ)-type entangled state
\begin{equation}
|\psi_2\rangle=(|g\rangle^{\otimes N}+|e\rangle^{\otimes N})/\sqrt{2}.
\end{equation}
Second, the atoms experience a free evolution governed by the atomic Hamiltonian $\hat{H}_{0}=\Delta\sum_j\hat{\sigma}_j^\dag\hat{\sigma}_j$ to encode $\omega_0$ into the probe state, which converts $|\psi_2\rangle$ into
\begin{equation}
|\psi(t)\rangle=(|g\rangle^{\otimes N}+e^{-iN\Delta t}|e\rangle^{\otimes N})/\sqrt{2}.\label{evlghz}
\end{equation}
Third, another CNOT gate is applied again to convert $|\psi(t)\rangle$ into $|\psi'(t)\rangle=\frac{|g\rangle+e^{-iN\Delta t}|e\rangle}{\sqrt{2}}\otimes |g\rangle^{\otimes (N-1)}$. Switching on the $\pi/2$ pulse again, $|\psi'(t)\rangle$ becomes
\begin{equation}
|\psi_\text{out}\rangle= [\cos(N\Delta t/ 2)|g\rangle+i\sin(N\Delta t/ 2)|e\rangle]\otimes |g\rangle^{\otimes N-1}.
\end{equation}
Measuring the excited-state population operator $\hat{O}=\hat{\sigma}_1^\dag\hat{\sigma}_1$ of the first atom, we obtain the results 1 with probability $P_1=\sin^2{N\Delta t\over 2}$ and 0 with probability $P_0=\cos^2{N\Delta t\over 2}$. They lead to $\langle \hat{O}\rangle=\sin^2{N\Delta t\over 2}$, $\Delta O=|\sin(N\Delta t)|/2$, and $F_{\omega_0} = N^2 t^2$. Repeating this experiment within a time duration $T$ yields $\upsilon=T/t$ sets of experimental results. Thus, according to the central limit theorem, the uncertainty of $\hat{O}$ is
\begin{equation}
\delta O=\frac{\Delta O}{\sqrt{T/t}}=\frac{|\sin (N\Delta t)|}{2}\sqrt{\frac{t}{T}}.
\end{equation}
The error propagation formula results in
\begin{equation}\label{RIdeal}
\delta \omega_0=(N^2Tt)^{-1/2},
\end{equation}
which is called the HL~\cite{Fink_2019,DeLeonardis2020}. Equation \eqref{RIdeal} also equals the sensitivity evaluated from the Cram\'{e}r-Rao bound \eqref{eq:eq2-1}.  Obviously, by using entanglement, the HL has an $N^{1/2}$-times enhancement over the SNL in Eq. \eqref{rmssnl}. The QFI of Eq. \eqref{evlghz} is $\mathcal{F}_{\omega_0}(t)=N^2t^2$, from which the ultimate sensitivity of $\omega_0$ reads
\begin{equation}
\delta\omega_0=(\upsilon \mathcal{F}_{\omega_0})^{-1/2}=(N^2Tt)^{-1/2}.
\end{equation}
It matches with Eq. \eqref{RIdeal}. Therefore, the measurement scheme used above saturates the quantum Cram\'{e}r-Rao bound and is optimal.
\subsection{Mach-Zehnder interferometer}\label{sbmzid}

\begin{figure}[tbp]
	\centering
	\includegraphics[scale=1.1]{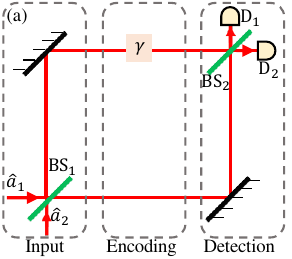}
	\caption{ Scheme of a Mach-Zehnder-interferometer based quantum metrology. Two fields interact at beam splitter BS$_1$ and propagate along two arms. One of the fields couples to a system with the potential influence of quantum noise, by which the estimated parameter $\gamma$ is encoded. After interfering at BS$_2$, the fields are detected by the detectors D$_1$ and D$_2$. Reproduced with permission \cite{PhysRevLett.123.040402}. Copyright 2019, American Physical Society.}\label{1f}
\end{figure}

A wide class of quantum metrology using quantized lights as probes is based on the Mach-Zehnder interferometer~\cite{Giovannetti2011,PhysRevD.23.1693,PhysRevLett.123.040402,PhysRevLett.104.103602}. In order to measure a frequency parameter $\gamma$ of a system, we choose two modes optical fields with
frequency $\omega_0$ as the probe. The encoding of $\gamma$ is realized by the time evolution $\hat{U}_0(\gamma,t)=\exp(-{i\hat{H}_0t})$ with
\begin{equation}
\hat{H}_0=\omega_0\sum_{m=1,2}\hat{a}_m^\dag\hat{a}_m+\gamma\hat{a}_2^\dag\hat{a}_2,
\end{equation}where the first term is the Hamiltonian of the fields and the second one is the linear interaction of the second field with the system \cite{PhysRevLett.107.083601,PhysRevLett.101.040403,PhysRevA.77.012317}. The evolution $\hat{U}_0(\gamma,t)$ accumulates a phase difference $\gamma t$ to the two fields, which is measured by the Mach-Zehnder interferometer. It has two beam splitters $\text{BS}_i$ ($i=1,2$) separated by the phase shifter $\hat{U}_0(\gamma,t)$ and two detectors $\text{D}_i$ (see Fig. \ref{1f}). Its input-output relation reads
\begin{equation}
|\Psi_\text{out}\rangle=\hat{V}\hat{U}_0(\gamma,t)\hat{V}|\Psi_\text{in}\rangle,\label{inoutre}
\end{equation} where $\hat{V}=\exp[i\frac{\pi}{4}(\hat{a}_{1}^{\dagger}\hat{a}_{2}+\hat{a}_{2}^{\dagger}\hat{a}_{1})]$ is the action of $\text{BS}_i$ \cite{PhysRevLett.104.103602}. We are interested in the quantum superiority of $|\Psi_\text{in}\rangle$ in metrology subject to an explicit measurement scheme. Thus we consider Caves's original scheme \cite{PhysRevD.23.1693}; i.e., the photon difference $\hat{M}=\hat{a}^\dag_1\hat{a}_1-\hat{a}^\dag_2\hat{a}_2$ is measured by $\text{D}_i$, which is also the most general measurement in the Mach-Zehnder interferometer. We can evaluate $\langle \hat{M}\rangle=\langle\Psi_\text{out}|\hat{M}|\Psi_\text{out}\rangle$ and $\delta ^2M=\langle\hat{M}^2\rangle-\langle \hat{M}\rangle^2$. The metrology sensitivity is obtained via
\begin{equation}
\delta\gamma={\delta M\over |\partial_\gamma\langle\hat{M}\rangle|}.
\end{equation}

If the input state is a product state of a coherent state and a vacuum state, i.e., $|\Psi_\text{in}\rangle=\hat{D}_{\hat{a}_1}|0,0\rangle$, with $\hat{D}_{\hat{a}}=\exp{(\alpha\hat{a}^{\dagger}-\alpha^{\ast}\hat{a})}$, which contains a mean photon number $N=|\alpha|^2$, then we have $\langle\hat{M}\rangle=N\cos(\gamma t)$ and $\langle\hat{M}^2\rangle=N^2\cos^2(\gamma t)+N$. It can be readily evaluated that
\begin{equation}
\text{min}\delta \gamma=1/(t\sqrt{N})\label{mzicvt}
\end{equation}when $\gamma t=m\pi/2$, with $m$ being a positive integer. In this case, the QFI reads $\mathcal{F}_\gamma=Nt^2$. The coincidence of the metrology sensitivity from $\mathcal{F}_\gamma$ with Eq. \eqref{mzicvt} indicates that the $\hat{M}$-measurement is optimal.

If the input state is a product state of a squeezed vacuum state and a coherent state, i.e., $|\Psi_{\text{in}}\rangle=\hat{D}_{\hat{a}_1}\hat{S}_{\hat{a}_2}|0,0\rangle$, with $\hat{S}_{\hat{a}}=\exp[(\xi^*\hat{a}^2-\xi\hat{a}^{\dagger 2}/2)]$ and $\xi=re^{i\phi}$, its total photon number is $N=|\alpha|^2+\sinh^2r$, which contains the ratio $\iota\equiv \sinh^2 r/N$ from the squeezed mode and can be regarded as the quantum resource of the scheme. To $|\Psi_{\text{out}}\rangle$, we have $\langle\hat{M}\rangle=[\sinh^2r-|\alpha|^2]\cos(\gamma t)$ and
\begin{eqnarray}
  \delta M &=&\{\cos ^{2}(\gamma t)[\left\vert \alpha \right\vert ^{2}+2\sinh^{2}r\cosh ^{2}r]+\sin ^{2}(\gamma t)\nonumber\\
  &&\times[\left\vert \alpha \cosh r-\alpha ^{\ast}\sinh re^{i\phi }\right\vert ^{2}+\sinh ^{2}r]\}^{\frac{1}{2}}.
\end{eqnarray}
Then the best precision of estimating $\gamma$ is obtained as
\begin{equation}
\min\delta\gamma=\frac{[(1-\iota)e^{-2r}+\iota]^{\frac{1}{2}}}{t\sqrt{N}|1-2\iota|}
\end{equation}
when $\phi=2\varphi$ and $\gamma t=(2m+1)\pi/2$ for $m\in \mathbb{Z}$. If the squeezing is absent, then $\min\delta\gamma|_{\iota=0}=(tN_0^{1/2})^{-1}$ is just the SNL. For $\iota\neq0$, using $e^{-2r}\simeq1/(4\sinh^2r)$ for $N\gg1$ and optimizing $\iota$, we have
\begin{equation}\label{ZL}
	\min\delta\gamma|_{\iota=(2\sqrt{N})^{-1}}=(tN^{3/4})^{-1},
\end{equation}
which is the Zeno limit~\cite{PhysRevLett.109.233601,doi:10.1063/1.5066028,PhysRevApplied.14.034065}. It beats the SNL and manifests the superiority of squeezing in metrology. The QFI of $|\Psi_\text{out}\rangle$ in this case is derived to be
\begin{equation}
\mathcal{F}_\gamma=t^2N\iota [4N(1-\iota)+1].
\end{equation}
Optimizing $\iota$, we obtain the best sensitivity $\min\delta\gamma|_{\iota=1/2}=(tN)^{-1}$, which is smaller than the Zeno limit in Eq. \eqref{ZL}. This means that the above measurement scheme is not optimal. Reference \cite{PhysRevLett.100.073601} revealed that the measurement of the photon numbers of both of the two output ports can saturate the quantum Cram\'{e}r-Rao bound.
Benefiting from the quantum-enhanced sensitivity, the squeezing has been used in gravitational-wave observatory~\cite{PhysRevLett.110.181101,PhysRevLett.123.231107,PhysRevLett.123.231108}.

If the input state is two-mode squeezed vacuum state $|\Psi_\text{in}\rangle=\hat{S}|0,0\rangle$, with $\hat{S}=e^{r(\hat{a}_1\hat{a}_2-\hat{a}_1^\dag\hat{a}_2^\dag)}$, whose total photon number is $N=2\sinh^2r$. In this case, we choose to measure the parity operator $\hat{\Pi}=e^{i\pi\hat{a}^\dag_1\hat{a}_1}$. Substituting $|\Psi_\text{in}\rangle$ into Eq. \eqref{inoutre}, we calculate $\langle\hat{\Pi}\rangle=\langle\Psi_{\text{out}}|\hat{\Pi}|\Psi_{\text{out}}\rangle=[1+N(2+N)\cos^{2}(\gamma t)]^{-1/2}$ and $\delta\Pi=(1-\langle\hat{\Pi}\rangle^{2})^{1/2}$, where $\hat{\Pi}^{2}=1$ has been used. Then the metrology sensitivity is evaluated via the error propagation formula $\delta\gamma=\frac{\delta\Pi}{|\partial_\gamma\langle\hat{\Pi}\rangle|}$ as
\begin{equation}
\min\delta\gamma=\big[2t\sqrt{N(2+N)}\big]^{-1},\label{idelsg}
\end{equation}
when $\gamma t=(2n+1)\pi/2$ with $n\in\mathbb{Z}$. It is remarkable to find that the best sensitivity is even smaller than the HL $\Delta\gamma\propto(tN)^{-1}$, which reflects the quantum superiority of the used squeezing and measured observable. The QFI of $|\Psi_\text{out}\rangle$ equals to
\begin{equation}
    \mathcal{F}_{\gamma}=t^2N(N+2)\sin^2(\gamma t).
\end{equation}The sensitivity obtained from $\mathcal{F}_\gamma$ matches well with Eq. \eqref{idelsg}. It verifies that the parity-measurement scheme saturates the quantum  Cram\'{e}r-Rao bound and is optimal.
We call such a sensitivity surpassing the HL the super-HL \cite{PhysRevLett.101.040403,Napolitano2011,PhysRevX.8.021022,PhysRevLett.126.070503}.

\subsection{Sagnac interferometer}
High-performance gyroscopes for rotation sensing are of pivotal significance for navigation in many types of air, ground, marine, and space applications. Based on the Sagnac effect, i.e., two counter-propagating waves in a rotating loop accumulate a rotation-dependent phase difference, gyroscopes have been realized in optical systems \cite{Khial2018,Lai2020,Srivastava2016,PhysRevLett.125.033605,PhysRevResearch.2.032069,Sanders:21}. The records for precision and stability of commercial gyroscopes are held by optical gyroscopes \cite{Culshaw_2005,LEFEVRE2014851}. However, their precision, which is proportional to the surface area enclosed by the optical path \cite{Arditty:81}, is still limited by the classical SNL. It dramatically constrains their practical application and further performance improvement. A quantum gyroscope based on the Sagnac interferometer can be established as follows.

\begin{figure}[tbp]
		\centering
		\includegraphics[width=.5\columnwidth]{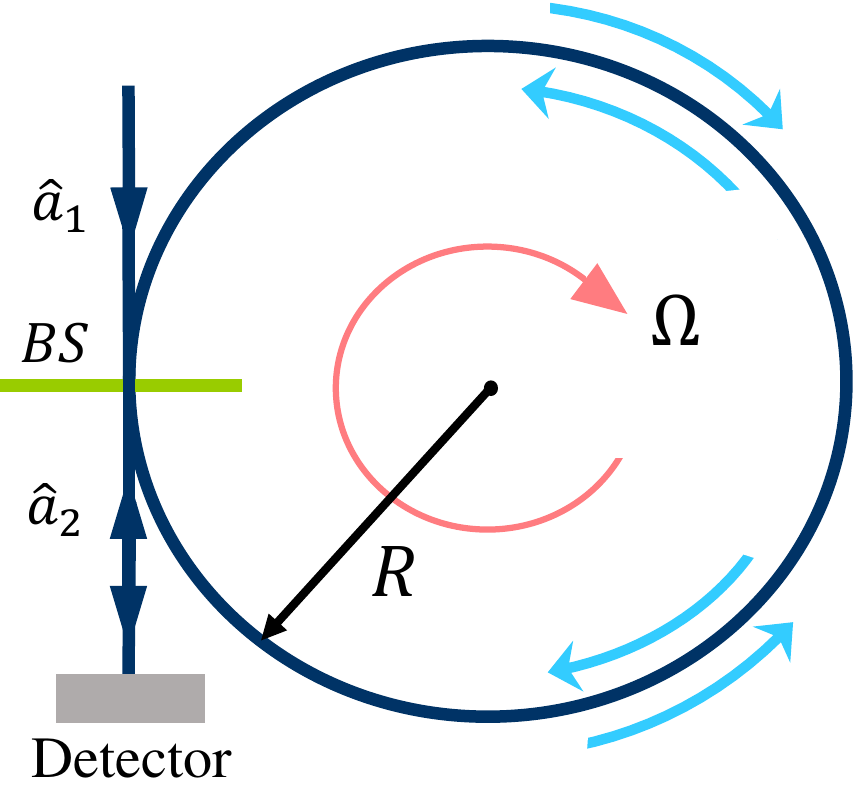}
	\caption{Schematic diagram of a quantum gyroscope based on the Sagnac interferometer. Reproduced with permission \cite{Jiao:23}. Copyright 2023,
Chinese Laser Press.	}
	\label{scmgr}
\end{figure}

\begin{table*}[t]
\begin{center}
\caption{Comparison of different scaling relations of the metrology error $\delta x$ of a physical quantity $x$ with respect to the number of quantum resources contained in different input states in the metrology schemes based on Mach-Zehnder interferometer and Ramsey spectroscopy, respectively.}\label{table:table1}
\setlength{\tabcolsep}{7.5pt}
\begin{tabular}{ccc}
  \hline
  \hline
    & Mach-Zehnder interferometer & Ramsey spectroscopy \\ \hline
  Coherent state & $\delta^{2}x\propto N^{-1}$ & -- \\
  NOON state & $\delta^{2}x\propto N^{-2}$~\cite{PhysRevLett.99.070801} & --\\
  Coherent state + Squeezed state & $\delta^{2}x\propto N^{-2}$~\cite{PhysRevLett.100.073601} & -- \\
  Two-mode squeezed state & $\delta^{2}x\propto(N^{2}+2N)^{-1}$~\cite{PhysRevLett.104.103602} & --\\
  Coherent spin state & -- & $\delta^{2}x\propto N^{-1}$ \\
  One-axis twisted state & -- & $\delta^{2}x\propto N^{-5/3}$~\cite{PhysRevA.47.5138} \\
  Two-axis twisted state & -- & $\delta^{2}x\propto N^{-2}$~\cite{PhysRevA.47.5138} \\
  GHZ state & -- & $\delta^{2}x\propto N^{-2}$~\cite{PhysRevLett.79.3865}  \\
  \hline
  \hline
\end{tabular}
\end{center}
\end{table*}
We choose two beams of quantized optical fields as the quantum probe. They propagating in opposite directions are input into a 50:50 beam splitter and split into clockwise and counter-clockwise propagating beams (see Fig. \ref{scmgr}). The setup rotates with an angular velocity $\Omega$ about the axis perpendicular to its plane. Thus the two beams accumulate a phase difference $\Delta \theta={\mathcal{N}4\pi k R^2 \Omega/c}$ when they re-encounter the beam splitter after $\mathcal{N}$ rounds of propagation in the circular path \cite{scully1997quantum}. Here $k$ is the wave vector, $c$ is the speed of light, and $R$ is the radius of the quantum gyroscope. Remembering the standing-wave condition $kR=n$~($n\in\mathbb{Z}$) of the optical fields propagating along the circular path and defining $\Delta t\equiv \mathcal{N}2\pi R/c$, we have $\Delta\omega\equiv \Delta\theta/\Delta t=2n\Omega$. Therefore, the quantum gyroscope can be equivalently treated as two counter-propagating optical fields with a frequency difference $\Delta\omega$ along the circular path. For concreteness, we choose the basic mode $n=1$. Then the optical fields in the quantum gyroscope can be quantum mechanically described by \cite{PhysRevA.95.012326}
\begin{equation}\label{ss3}
\hat{H}_{S}=\omega_{0}\sum_{l=1,2}\hat{a}_{l}^{\dagger}\hat{a}_{l}+\Omega(\hat{a}_{1}^{\dagger}\hat{a}_{1}-\hat{a}_{2}^{\dagger}\hat{a}_{2}),
\end{equation}where $\hat{a}_l$ is the annihilation operator of the $l$th field with frequency $\omega_0$. The optical fields couple to the beam splitter twice and output in the state $|\Psi_{\text{out}}\rangle =\hat{V}\hat{U}_{0}(\Omega,t)\hat{V}|\Psi_{\text{in}}\rangle $, where $\hat{U}_{0}(\Omega,t)=\exp(-i\hat{H}_{S}t)$ is the evolution operator of the fields and $\hat{V}=\exp[i\frac{\pi}{4}(\hat{a}_{1}^{\dagger}\hat{a}_{2}+\hat{a}_{2}^{\dagger}\hat{a}_{1})]$ describes the action of the beam splitter. Thus the angular velocity $\Omega$ is encoded into the state $|\Psi_{\text{out}}\rangle$ of the optical probe via the unitary evolution.

To exhibit the quantum superiority, we employ two-mode squeezed vacuum state as the input state $|\Psi_{\text{in}}\rangle=\hat{S}|0,0\rangle$, where $\hat{S}=\exp[r(\hat{a}_{1}\hat{a}_{2}-\hat{a}_{1}^{\dagger}\hat{a}_{2}^{\dagger})]$ is the squeeze operator. The total photon number of this input state is $N=2\sinh^{2}r$, which is the quantum resource of our scheme. The parity operator $\hat{\Pi}=\exp(i\pi\hat{a}_{1}^{\dagger}\hat{a}_{1})$ is measured at the output port \cite{PhysRevLett.104.103602}. Similar to the last example in Sec. \ref{sbmzid}, we have
\begin{equation}
\min\delta\Omega=\big[2t\sqrt{N(2+N)}\big]^{-1},\label{idel}
\end{equation}
when $\Omega t=(2n+1)\pi/4$ with $n\in\mathbb{Z}$. Super-HL sensing to the angular velocity is achieved. It can be verified that this measurement scheme saturates the quantum Cram\'{e}r-Rao bound governed by the QFI.

We summarize the comparison of different scaling relations of the metrology error $\delta x$ of a physical quantity $x$ with respect to the number of quantum resources contained in different input states in the metrology schemes based on Mach-Zehnder interferometer and Ramsey spectroscopy in Table \ref{table:table1}.

\section{Other quantum resources}\label{otqsr}

\subsection{Spin Squeezing}
Spin-squeezed states are a class of collective-spin states having squeezed spin variance along a certain direction, at the cost of anti-squeezed variance along an orthogonal direction~\cite{PhysRevA.46.R6797,RevModPhys.90.035005,PhysRevA.47.5138,PhysRevA.50.67}. Spin squeezing is one of the most successful quantum resources that can witness large-scale quantum-favored beating to the SNL in the NISQ era.

Consider an ensemble of $N$ two-level atoms or spin-$1/2$ particles in a quantum state $|\Psi\rangle$. Its collective spin operator is defined as $\hat{\bf J}=\sum_{j=1}^N\hat{\pmb\sigma}_j/2$. The mean-spin direction is ${\bf n}_0=\langle \hat{\bf J}\rangle/|\langle \hat{\bf J}\rangle|$, with $\langle \hat{\bf J}\rangle=\langle\Psi|\hat{\bf J}|\Psi\rangle$. To properly use this kind of collective spin state, we design a generalized Ramsey spectroscopy as follows \cite{PhysRevA.50.67}. First, we perform a rotation with angle $\vartheta$ along the axis ${\pmb \eta}$ such that ${\pmb \eta}\cdot{\bf n}_0=0$. Then, we estimate $\vartheta$ by measuring the operator $\hat{J}_\perp={\pmb\alpha}\cdot\hat{\bf J}$, where ${\pmb\alpha}\cdot{\bf n}_0={\pmb\alpha}\cdot {\pmb \eta}={\bf n}_0\cdot {\pmb \eta}=0 $. In the Heisenberg picture, the rotation converts the measured operator $\hat{J}_\perp$ into
\begin{equation}
\hat{J}^\text{out}_\perp=e^{i\vartheta\hat{J}_\eta}\hat{J}_\perp e^{-i\vartheta\hat{J}_\eta}=\cos\vartheta \hat{J}_\perp-\sin\vartheta \hat{J}_{n_0}.
\end{equation}
It follows that $\langle\hat{J}^\text{out}_\perp\rangle= -\sin\vartheta \langle\hat{J}_{n_0}\rangle$ and
\begin{eqnarray}
\delta^2 J^\text{out}_\perp&=&\cos^2\vartheta \delta^2 J_\perp+\sin^2\vartheta\delta ^2 \hat{J}_{n_0}\nonumber\\
&&-{\sin (2\vartheta)\over 2}\langle[\hat{J}_\perp,\hat{J}_{n_0}]_+\rangle.
\end{eqnarray}The best phase sensitivity is calculated via the error propagation formula as
\begin{equation}
\delta\vartheta=\min_\vartheta{\delta J^\text{out}_\perp\over |\cos\vartheta\langle\hat{J}_{n_0}\rangle |}={\delta J_\perp\over | \langle\hat{\bf J}\rangle |},
\end{equation}where $|\langle\hat{\bf J}\rangle |=|\langle \hat{J}_{n_0}\rangle{\bf n}_0|=|\langle \hat{J}_{n_0}\rangle|$ has been used.

Consider the phase sensitivity obtained in the generalized Ramsey spectroscopy by using the so-called spin-coherent state defined as $|\theta,\phi\rangle={e^{\zeta\hat{J}_-}\over (1+|\zeta|^2)^j}|j,j\rangle$, where $\zeta=e^{i\phi}\tan{\theta\over 2}$ and $|j,j\rangle$ is the common eigen state of $|\hat{J}^2,\hat{J}_z\rangle$ with $j=N/2$. For this state, the mean-spin direction is ${\bf n}_0=(\sin\theta\cos\phi,\sin\theta\sin\phi,\cos\theta)$ and the expectation $|\langle\hat{\bf J}\rangle|=N/2$. After some algebra, the variance $\delta J_\perp=\sqrt{N}/2$ is evaluated. Therefore, we have
\begin{equation}
    \delta\vartheta_\text{scs}=N^{-1/2},
\end{equation}which is just the SNL. The squeezing parameter of a spin-squeezed state is defined as the ratio of the phase sensitivities obtained via the spin-squeezed state and the spin-coherent state \cite{PhysRevA.46.R6797}, i.e.,
\begin{equation}
    \xi_R\equiv{\delta\vartheta\over \delta\vartheta_\text{scs}}=\sqrt{N}{(\delta J_\perp)_\text{min}\over | \langle\hat{\bf J}\rangle |}.\label{wldsq}
\end{equation}
When $\xi_R<1$, the state is said to be spin squeezed, by which the obtained metrology sensitivity of measuring $\vartheta$ in the generalized Ramsey spectroscopy beats the SNL $\xi_R$ times. This requires that the state has a minimal spin fluctuation in the plane orthogonal to the mean-spin direction smaller than the one of the spin coherent state, i.e., $\min\delta\hat{J}_{\perp}<\sqrt{N}/2$. $\xi_R$ gives the quantum gain of the frequency sensitivity in the generalized Ramsey spectroscopy obtained by the spin-squeezed state over the SNL obtained by the spin coherent state. It has been revealed that a collective-spin state with spin squeezing contains genuine many-body entanglement \cite{PhysRevLett.112.155304}, which is the intrinsic reason for the superiority of the spin squeezed state in enhancing metrology sensitivity.

First, we consider the superposition of two Dicke states \cite{PhysRevA.50.67}
\begin{equation}
    |\Psi_\text{SDS}\rangle=(|j,{1/ 2}\rangle+|j,-{1/ 2}\rangle)/\sqrt{2},
\end{equation}where $j=N/2$ with $N$ being odd. One can evaluated $\langle\hat{\bf J}\rangle=({j+1/2\over 2},0,0)$. Thus $\hat{J}_\perp=\cos x\hat{J}_y+\sin x\hat{J}_z$, with $x\in[0,2\pi]$. It is straightforward to derive that
\begin{eqnarray}
\min_x\delta^2 J_\perp&=&\{\langle (\hat{J}_y^2+\hat{J}_z^2)\rangle-[\langle (\hat{J}_y^2-\hat{J}_z^2)\rangle^2\nonumber\\&&+\langle [\hat{J}_y,\hat{J}_z]_+\rangle^2]^{1/2}\}/2.\label{jpert}
\end{eqnarray}
According to $\langle [\hat{J}_y,\hat{J}_z]_+\rangle=0$ for $|\Psi_\text{SDS}\rangle$, we have $(\delta^2 J_\perp)_\text{min}=\langle\hat{J}_z^2\rangle=1/4$.  Substituting it into Eq. \eqref{wldsq}, we obtain
\begin{equation}
    (\xi_R)_\text{SDS}=2\sqrt{N}/( N+1)\simeq 2N^{-1/2}.
\end{equation}
It indicates that the HL of the frequency sensitivity could be obtained if $|\Psi_\text{SDS}\rangle$ is used in the generalized Ramsey spectroscopy.

Second, we consider the one-axis twisted state defined as \cite{PhysRevA.47.5138}
\begin{equation}
	|\Psi_{\text{OAT}}\rangle =e^{-i\mu \hat{J}_{z}^{2}/2}|\pi/2,0\rangle.
\end{equation}
The expectation value of $\hat{\bf J}$ is $\langle\hat{\bf J}\rangle=(j\cos^{2j-1}(\mu/2),0,0)$. Thus $(\delta{J}_\perp)_\text{min}$ has the same form as Eq. \eqref{jpert}. Using $\langle\hat{J}_z^2\rangle=j/2$ and
\begin{eqnarray}
    \langle\hat{J}_y^2\rangle={j\over2}[j+{1\over 2}-(j-{1\over 2})\cos^{2j-2}\mu],\\
    \langle[\hat{J}_y,\hat{J}_z]_+\rangle=j(2j-1)\cos^{2j-2}(\mu/2)\sin(\mu/2),
\end{eqnarray} we obtain
\begin{equation}
    \min_x \delta^2J_\perp={j\over 2}[1+{j-1/2\over 2}(A-\sqrt{A^2+B^2})],\label{ssdv}
\end{equation}where $A=1-\cos^{2j-2}\mu$ and $B=4\cos^{2j-2}(\mu/2)\sin(\mu/2)$. Under the condition $j\gg 1$ and $\mu\ll 1$, Eq. \eqref{ssdv} is approximated as $\min_x \delta^2J_\perp={j\over 2}({1\over 4\alpha^2}+{2\beta^2\over 3})$, where $\alpha=j\mu/2$ and $\beta=j\mu^2/4$. It reaches its minimum $(\delta J_\perp)_\text{min}={1\over \sqrt{2}}({j\over3})^{1/6}$ when $\mu=24^{1/6}j^{-2/3}$. Substituting it into Eq. \eqref{wldsq}, we obtain
\begin{equation}
    (\xi_R)_\text{OAT}\propto j^{-1/3}\propto N^{-1/3}.
\end{equation}
Thus a metrology sensitivity $N^{-5/6}$ would be achieved if $|\Psi_{\text{OAT}}\rangle$ is used in the generalized Ramsey spectroscopy.

Third, we consider the two-axis twisting state defined as
\begin{equation}
	|\Psi_{\text{TAT}}\rangle =e^{-\theta(\hat{J}_{+}^{2}-\hat{J}_{-}^{2})/2}|j,-j\rangle ^{\otimes N},
\end{equation}
where the mean-spin direction is along the $z$-axis. Unfortunately, the two-axis twisting model cannot be solved analytically for arbitrary $N$. The numerical calculations reveal that the spin-squeezing parameter $\xi_R$ scales with the atom number $N$ as $(\xi_R)_\text{TAT}\sim N^{-1/2}$, which leads to a sensitivity scaling as the HL $\delta\vartheta\sim N^{-1}$.

Spin-squeezed states reflect the absolute superiority of the entanglement-enhanced sensitivity in quantum metrology. The distinctive role of atomic spin squeezing in improving the sensitivity makes it valuable for the applications in quantum gyroscope \cite{PhysRevLett.98.030407,PhysRevLett.114.063002}, atomic clocks \cite{PhysRevLett.104.250801,PhysRevLett.112.190403,Komar2014,PhysRevLett.117.143004,PedrozoPenafiel2020a}, magnetometers \cite{PhysRevLett.104.133601,PhysRevLett.109.253605,PhysRevLett.113.103004,Bao2020}, and gravimetry \cite{PhysRevLett.125.100402}.

\subsection{Quantum criticality}\label{dfdsa2}

Quantum criticality can also act as a resource to enhance the sensitivity in metrology \cite{RevModPhys.90.035006,Garbe_2022}. It takes advantage of criticality associated with continuous quantum phase transitions, in which the energy gap above the ground state closes in the thermodynamic limit, to realize high-precision measurements to physical quantities. Consider a Hamiltonian $\hat{H}(g)$ with the decomposition $\hat{H}(g)=\sum_{n}E_{n}(g)|\psi_{n}(g)\rangle\langle\psi_{n}(g)|$, the QFI with respect to the parameter $g$ that drives the quantum phase transition for the ground state $|\psi_{0}(g)\rangle$ reads \cite{Gietka2021adiabaticcritical,PhysRevE.76.022101}
\begin{equation}
\mathcal{F}_{g}=4\sum_{n\ne0}\frac{|\langle\psi_{n}(g)|\partial_{g}\hat{H}(g)|\psi_{0}(g)\rangle|^{2}}{[E_{n}(g)-E_{0}(g)]^{2}}. \label{QFI critical}
\end{equation}
It is clear that, if the energy gap above the ground state closes, the QFI diverges due to the vanishing denominator. This property leads to an arbitrarily high estimation precision in the thermodynamic limit.
In general, the lowest excitation energy tends to zero as $(E_1-E_0)\sim\xi^{-z}$ near the critical point $g_{c}$, where $z$ is the dynamical critical exponent and $\xi\sim\Lambda^{-1}|g-g_{c}|^{-\nu}$ is the correlation length. Here $\Lambda$ is the momentum cutoff determined by the inverse lattice spacing and $\nu$ is the critical exponent \cite{sachdev_2000}. Combined with Eq. \eqref{QFI critical}, the QFI diverges as $\mathcal{F}_{g}\sim |g-g_{c}|^{-2z\nu}$ near the critical point. A commonly used quantum critical metrology protocol is to adiabatically prepare a ground state with the parameters being near the critical point, then measure the observable to estimate the quantity that drives the occurrence of the quantum phase transition.

First, we consider applying the one-dimensional quantum transverse-field Ising model \cite{PhysRevA.78.042106,PhysRevLett.121.020402} in quantum metrology. Its Hamiltonian reads
\begin{equation}
\hat{H}=-J\sum_{i=1}^{L-1}\hat{\sigma}_{i}^{x}\hat{\sigma}_{i+1}^{x}-h\sum_{i=1}^{L}\hat{\sigma}_{i}^{z},\label{qimd}
\end{equation}
where $\hat{\sigma}_{i}^{x/z}$ are Pauli operators and $L$ is the size of spin chain. Using the standard Jordan-Wigner, Fourier, and Bogoliubov transformations, Eq. \eqref{qimd} is rewritten as $\hat{H}=\sum_{k>0}\Lambda_{k}(\hat{\eta}_{k}^{\dagger}\hat{\eta}_{k}-1)$, where $k=(2n+1)\pi/L$, with $n=0,\cdots,L/2-1$, $\hat{\eta}_{k}$ is the fermion annihilation operator, and $\Lambda_{k}=\sqrt{(J\cos k+h)^{2}+(J\sin k)^{2}}$ is the excitation energy. It has a quantum phase transition from an ordered phase to a paramagnetic phase at the critical point $h=J$. The QFI for the parameter $J$ in the ground state is
\begin{equation}
\mathcal{F}_{J}=\sum_{k}\frac{(J+\frac{z_c}{L})^{2}\sin^{2}k}{[\frac{z_c^{2}}{L^{2}}+4J(J+\frac{z_c}{L})\cos^{2}{k\over 2}]^{2}},
\end{equation}where $z_c\equiv L(h-J)$ is the scaling variable.
Since $\partial_{z_c} \mathcal{F}_J |_{z_c\rightarrow0}=0$, $\mathcal{F}_{J}$ has a maximum at $z_c=0$ for all values of $L$. It scales with $L$ as the HL, i.e., $\mathcal{F}_{J}\approx\frac{L^{2}}{8J^{2}}$ in the critical region, while it scales as $L^{1}$ in the off-critical region \cite{PhysRevLett.121.020402,PhysRevLett.99.100603}.

\begin{figure}[tbp]
\begin{center}
  \includegraphics[width=\columnwidth]{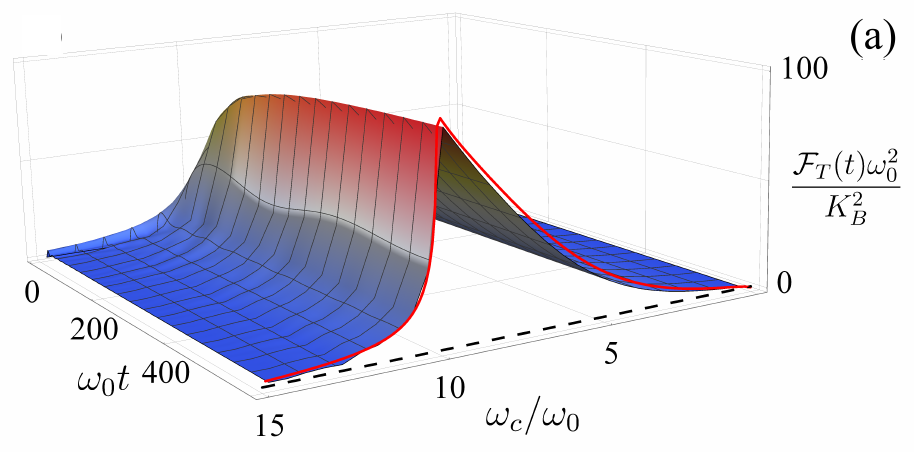}\\
  \vspace{.3cm}
  \includegraphics[width=.95\columnwidth]{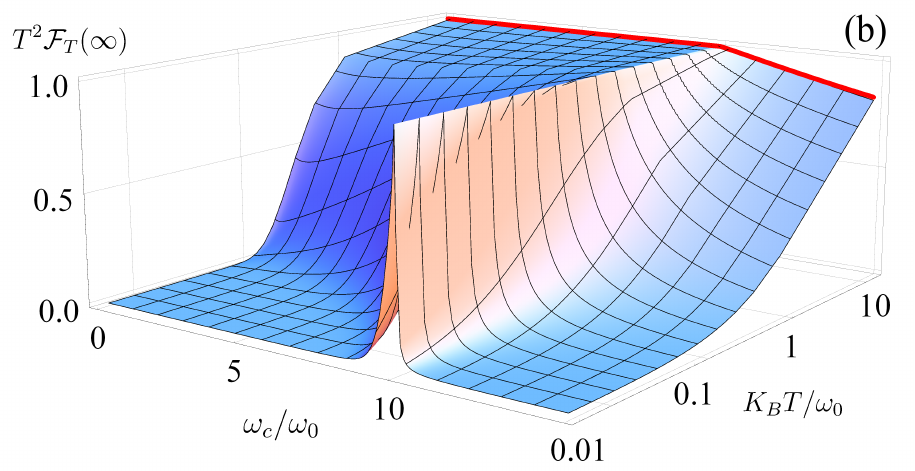}
  \caption{(a) Evolution of the non-Markovian QFI for different $\omega_c$. (b) Steady-state QFI for different $T$ and $\omega_c$. We use $\eta=0.1$, $s=1$, and $T=0.1\omega_0/K_B$. A quantum phase transition occurs at $\omega_c/\omega_0=10$. Reproduced
with permission \cite{PhysRevApplied.17.034073}. Copyright 2022, American Physical Society.}
  \label{timedep}
\end{center}
\end{figure}
Second, quantum critical metrology can be used to develop quantum thermometry. The precise measuring of temperature at the quantum level plays a more and more important role in the emerging fields of quantum thermodynamics and quantum technologies \cite{Mehboudi2019}. Although much effort has been made to enhance the precision of temperature measurement by using different quantum features, precisely measuring low temperature is still extremely challenging because the measured temperature errors in the existing quantum thermometry schemes are commonly divergent with decreasing temperature \cite{PhysRevLett.114.220405,PhysRevLett.119.090603,Pasquale2016,PhysRevLett.123.180602,PhysRevLett.125.080402}. Reference \cite{PhysRevApplied.17.034073} presents non-Markovian quantum thermometry to measure the temperature of a quantum reservoir. The reservoir is at equilibrium initially, i.e., $\rho_\text{R}(0)=\prod_{ k}e^{-\beta\omega_{k} \hat{b}^{\dagger}_{k} \hat{b}_{k}}/\text{Tr}[e^{-\beta\omega_{k} \hat{b}^{\dagger}_{k} \hat{b}_{k}}]$, where $\beta=(K_{B}T)^{-1}$ and $K_{B}$ is the Boltzmann constant. A continuous-variable system is used as the quantum thermometer. The encoding dynamics are governed by the Hamiltonian
\begin{equation}
	\hat{H}=\omega_{0}\hat{a}^{\dagger} \hat{a}+\sum_{k} [\omega_{k}\hat{b}^{\dagger}_{k} \hat{b}_{k} +g_{k} (\hat{a}^{\dagger} \hat{b}_{k}+\hat{b}^{\dagger}_{k}\hat{a})],\label{Hamlt}
\end{equation}
where $\hat{a}$ and $\hat{b}_{k}$ are the annihilation operators of the thermometer with frequency $\omega_{0}$ and the $k$th reservoir mode with frequency $\omega_{k}$ of the measured reservoir, and $g_k$ is their coupling strength. Their coupling is further characterized by the spectral density $J(\omega)=\sum_{k} g^{2}_{k} \delta (\omega-\omega_{k})$. The Ohmic-family spectral density $J(\omega)=\eta\omega^{s}\omega_{c}^{1-s} e^{-\omega/\omega_{c}}$, where $\eta$ is a dimensionless coupling constant, $\omega_{c}$ is a cutoff frequency, and $s$ is an Ohmicity index, is used. The ground state of Eq. \eqref{Hamlt} has a quantum phase transition at the critical point $\omega_{0}/\omega_{c}=\eta\underline{\Gamma}(s)$, where $\underline{\Gamma}(s)$ is Euler's $\gamma$ function. Taking the Ohmic spectral density as an example, Fig. \ref{timedep}(a) plots the non-Markovian evolution of $\mathcal{F}_T(t)$ for different cutoff frequencies $\omega_c$. It can be seen that $\mathcal{F}_T(t)$ gradually increases with time from zero to $\omega_c$-dependent stable values, which are larger than the Markovian approximate one $\bar{F}_T(\omega_0)$ in the full-parameter regime. Another interesting feature is that an obvious maximum of the QFI is present at $\omega_0=\eta\omega_c$, where the quantum phase transition occurs. Figure \ref{timedep}(b) shows the steady-state QFI $\mathcal{F}_T(\infty)$ for different $T$ and $\omega_c$. It clearly demonstrates that, at the critical point, the QFI scales with the temperature as $\mathcal{F}_T\sim T^{-2}$ in the full-temperature regime, which means that the performance of the non-Markovian quantum thermometry becomes better and better with a decrease of the temperature. This successfully solves the problem of the conventional schemes where the QFI tends to zero in the low-temperature regime. Besides the temperature of the reservoir, the criticality can also be used to sense the quantities in the spectral density \cite{PhysRevA.103.L010601,PhysRevApplied.15.054042}, which plays an important role in understanding the reservoir-induced decoherence to quantum systems.

The ``super-HL scaling'' in quantum critical metrology is another interesting topic. It was reported that the criticality can boost the QFI to $\mathcal{F}_{\lambda}\sim L^{2 \alpha}$ with $2<\alpha<3$ in Refs.  \cite{PhysRevB.77.245109,PhysRevB.78.115114,PhysRevB.88.195101,PhysRevA.101.043609,PhysRevResearch.5.013087,PRXQuantum.3.010354}. This contradicts with the result that the sensitivity would be bounded by the HL  $\mathcal{F}_{\lambda}\sim L^{2}$ if no $k$-order nonlinear terms, with $k\ge2$, are involved in the encoding Hamiltonian \cite{PhysRevLett.98.090401,Napolitano2011,PhysRevLett.126.070503,Yin2023,Tsarev_2019,NIE2018469,PhysRevLett.119.010403,XIE2022105957,PhysRevLett.118.140403,PhysRevLett.112.120405}. This contradiction was studied by considering the preparation time of the critical ground state \cite{PhysRevX.8.021022}. The ground state is generally adiabatically prepared, i.e., the initial ground state adiabatically evolves to the desired critical one. However, the energy gap near the critical point becomes smaller and smaller, and hence any change needs to be infinitely slow to keep the adiabaticity. This is known as the critical slowing down. Therefore, quantum-critical metrology inevitably requires a divergent preparation time. It was found that if the scaling of time is considered, the HL $\mathcal{F}_{\lambda}\sim L^{2}$ would be recovered and the ``super-HL'' in quantum critical metrology with local Hamiltonian would not present anymore \cite{PhysRevX.8.021022,PhysRevLett.124.120504,Gietka2021adiabaticcritical}.

To overcome the long time cost in adiabatic preparing of the critical ground state, a dynamic framework for critical metrology was proposed \cite{PhysRevLett.126.010502}. Consider a dynamical evolution state $|\psi_{\lambda}(t)\rangle=e^{-i\hat{H}_{\lambda}t}|\psi\rangle$ from an initial state $|\psi\rangle$, where $\hat{H}_{\lambda}$ is a family of Hamiltonians that all nearest energy level spacings have equal value $\sqrt{\Delta}$ near the critical point \cite{PhysRevA.90.022117, PhysRevLett.126.010502}. It has been proven that the QFI of estimating $\lambda$ scales as $\mathcal{F}_{\lambda}(t)\sim[\sin(\sqrt{\Delta}t)-\sqrt{\Delta}t]^{2}/\Delta^{3}$ \cite{PhysRevLett.126.010502}. Under the condition $\sqrt{\Delta}t\sim O(1)$, the QFI scales with $\Delta$ as $\mathcal{F}_{\lambda}\sim\Delta^{-2}t^{2}$ and is divergent near the critical point $\Delta\rightarrow0$. Due to the nonanalytical nature of the whole spectrum, this scaling behavior holds for any initial state. It efficiently avoids the time cost in adiabatically preparing the critical ground state. We take the quantum Rabi model as an example to illustrate this. Its Hamiltonian reads
\begin{equation}
\hat{H}_{\lambda}=\omega \hat{a}^{\dagger}\hat{a}+\frac{\omega_{0}}{2}\hat{\sigma}_{z}-\lambda\hat{\sigma}_{x}(\hat{a}+\hat{a}^{\dagger}).
\end{equation}
Setting $g\equiv 2\lambda/\sqrt{\omega\omega_{0}}$, this model has a critical point $g_{c}=1$, where a normal to superradiant phase transition occurs and the energy gap scales as $\varepsilon\sim|1-g^{2}|^{\nu z}$, with the critical exponent $\nu z=1/2$ \cite{PhysRevLett.115.180404,PhysRevLett.119.220601}. By defining $\Delta_{g}\equiv4(1-g^{2})$ and setting initial state $|\psi\rangle=|\downarrow\rangle\otimes |\psi\rangle_b$, the QFI for estimating $g$ with the evolved state is derived as
\begin{equation}
\mathcal{F}_{g}(t)\approx16g^{2}\frac{[\sin(\sqrt{\Delta_{g}}\omega t)-\sqrt{\Delta_{g}}\omega t]^{2}}{\Delta_{g}^{3}}\text{Var}[\hat{P}]_{|\psi\rangle_{b}},\label{eq:CQP-Rabi-QFI}
\end{equation}
with $\hat{P}=i(\hat{a}^\dag-\hat{a}^{\dagger})/\sqrt{2}$ and $|\psi\rangle_{b}$ being the bosonic part of initial state. It can be directly found that $\mathcal{F}_{g}(t)$ diverges when $g\rightarrow 1$. Notice that the evolution time needs to satisfy $\sqrt{\Delta_{g}}t\sim O(1)$, hence the dynamical protocol also requires a long evolution time near the critical point. But compared with the ground state critical quantum metrology, in which the adiabatic evolution time $t\sim1/\Delta_{g}$ \cite{PhysRevLett.124.120504,Garbe_2022}, such dynamical protocol greatly reduces the time cost.

In recent years, quantum critical metrology has been extended to many other transition types, such as topological phase transition \cite{PhysRevB.100.184417}, temperature driven phase transition \cite{PhysRevA.104.022612}, the first-order quantum phase transitions \cite{10.1063/1.5121558,Yang:19}, nonlinear critical model \cite{e24081015}, dynamical quantum phase transition \cite{PhysRevA.93.022103,PhysRevResearch.5.013087,PhysRevB.106.014313}, and Floquet induced criticality \cite{PhysRevLett.127.080504}.

\subsection{Quantum Chaos}

It was pointed out in Refs. \cite{Fiderer2018,PhysRevA.103.023309} that the quantum chaos can also be used to enhance the sensitivity of quantum metrology. Consider a collection of periodically kicked spins as the probe to detect a classical magnetic field. Its Hamiltonian reads
\begin{equation}
\hat{H}_{\text{KT}}(t)=\alpha \hat{J}_{z}+\frac{k}{2j+1}\hat{J}_{y}^{2}\sum_{n=-\infty}^{\infty}\tau\delta(t-n\tau),\label{pktp}
\end{equation}
where $\hat{J}_{y/z}$ are the collective spin operators, $j=N/2$ is the quantum number of $\hat{\mathbf{J}}^{2}$, and $\alpha$ and $k$ are the strengths of the magnetic field to be estimated and of the kicking field, respectively. The $\hat{J}_{y}^{2}$ term introduces the nonlinearity and leads to the chaotic dynamics when $k\ge3.0$ \cite{Chaudhury2009}. To use the chaotic resource instead of initial entanglement, the metrology protocol is as follows. First, we prepare the probe in a spin-coherent state $|\psi(0)\rangle=|\theta,\phi\rangle$, which is a product state of each spin state. Second, we encode $\alpha$ into the prove state via the evolution governed by Eq. \eqref{pktp}, i.e., $|\psi_\alpha(t)\rangle=\hat{\mathcal T}e^{-i\int_0^t\hat{H}_\text{KT}(\tau)d\tau}|\psi(0)\rangle$. Finally, we estimate the parameter $\alpha$ from $|\psi_{\alpha}(t)\rangle$. For comparison, when $k=0$, the system is totally integrable and such metrology protocol reduces to the standard Ramsey spectroscopy in Sec. \ref{rmsspc}.

\begin{figure}
\includegraphics[width=.45\columnwidth]{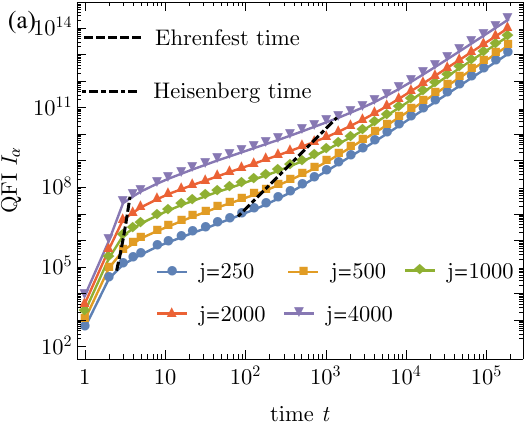}~~\includegraphics[width=.45\columnwidth]{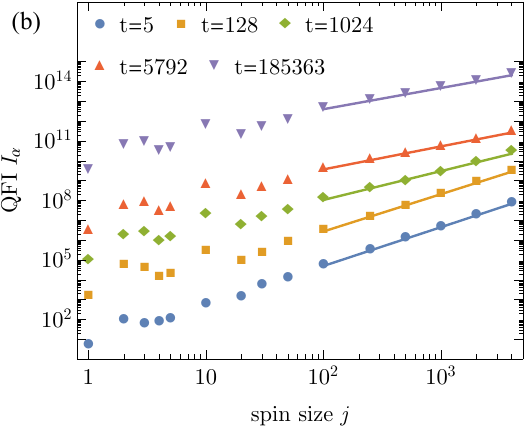}\caption{\label{fig:KickedTop} (a) $t$ scaling of the QFI of $\alpha$ $I_\alpha\equiv\mathcal{F}_{\alpha}$ in
the strongly chaotic case. Dashed and dash-dotted lines indicate Ehrenfest and Heisenberg times, respectively. (b) $j$ scaling of the QFI. Fits have slopes $1.96$, $1.88$, $1.46$, $1.16$, and $1.08$ in increasing order of $t$. Kicking strength $k=30$, and initial coherent state at $(\theta,\phi)=(\pi/2,\pi/2)$ in all plots. Reproduced with permission~\cite{Fiderer2018}. Copyright 2018, Springer Nature. }
\end{figure}

There are two important time scales in chaotic systems \cite{GORIN200633}. One is the Ehrenfest time $t_\text{E}={1\over\lambda}\ln{\Omega\over h^d}$, where $\lambda$ is the Lyapunov exponent, $\Omega$ and $h$ are the volumes of the phase space and a Planck cell, respectively, and $d$ is the number of the degrees of freedom. It is a time scale at which the correspondence between classical and quantum dynamics begins to break down. The other is the Heisenberg time $t_\text{H}=\hbar/\Delta$, where $\Delta$ is the mean energy level spacing. It is a time scale after which the dynamics resolves the discrete nature of the energy spectrum and a quasi-periodic behavior may be found. For Eq. \eqref{pktp} in the chaotic regime when $k=30$ under the initial condition $|\psi(0)\rangle=|\pi/2,\pi/2\rangle$, a detail analysis shows that $t_\text{E}=0.4\ln(2j+1)$ and $t_\text{H}=(2j+1)/6$.

It was found that
\begin{equation}
    \mathcal{F}_\alpha\propto \begin{cases}
tj^2, & t_\text{E}<t<t_\text{H} \\
t^2j, & t\gg t_\text{H}
\end{cases}.
\end{equation}
The numerical results are shown in Fig. \ref{fig:KickedTop}. Figure \ref{fig:KickedTop}(a) reveals that a transition from the $t$-scaling when $t\approx t_\text{E}$ to the $t^{2}$-scaling when $t\ge t_\text{H}$ occurs for different $j=N/2$. Figure \ref{fig:KickedTop}(b) confirms that the scaling relation of $\mathcal{F}_\alpha$ can go beyond the SNL when $t<t_\text{H}$ even without the initial entanglement. It reveals the sensitivity enhancement caused by the chaotic dynamics. A similar protocol that uses chaotic dynamics to estimate magnetic fields was discussed in Ref. \cite{PhysRevA.103.023309}. From the close relations between chaos, thermalization, and entanglement widely studied in Refs. \cite{Rigol,RevModPhys.91.021001,Greiner,Nandkishore}, a rough understanding of this enhancement is that the entanglement as the quantum resource is not prepared in the initial state, but is generated during the encoding dynamics governed by quantum chaos.

As a final remark in this section, quantum metrology schemes realizing the super-HL have gained popularity in recent years. It was found that, besides quantum criticality reviewed in Sec. \ref{dfdsa2}, squeezing~\cite{PhysRevLett.104.103602,PhysRevA.104.042609,PhysRevLett.69.3598,PhysRevA.82.045601,Jiao:23}, nonlinear effects~\cite{PhysRevLett.119.010403,PhysRevA.88.013817,PhysRevResearch.3.023222,PhysRevA.96.041801}, and time-dependent quantum control~\cite{PhysRevLett.98.090401,PhysRevLett.126.070503,PhysRevResearch.4.013133} also can make the metrology sensitivity outperform the HL and achieve a scaling relation $\delta\theta\sim N^{-\nu}$ with $\nu>1$. It is noted that a phase estimation error smaller than the inverse of the mean photon number was called the sub-HL in Refs. \cite{PhysRevLett.104.103602,PhysRevLett.108.210404,PhysRevA.88.060101}.

\section{Effects of Decoherence on Quantum Metrology}\label{effnqm}

From the foregoing sections, we see that quantum metrology works by using quantum effects, such as quantum coherence, squeezing, and entanglement, to develop revolutionary techniques for enhancing measurement precision. The main idea is to actively control the quantum states of relevant systems in the desired manner to realize more precise measurements of physical quantities than the achievable precision limit in classical physics. Although many exciting progresses have been made, quantum metrology is still in the proof-of-principle phase. It still has not shown its superiority in absolute sensitivities over its classical counterparts. This is because the practical realization of quantum metrology is challenged by different kinds of noise-induced decoherence in its stability and scalability. As a ubiquitous phenomenon in the microscopic world, decoherence is caused by the inevitable interaction of quantum systems with their environments. It makes the quantum resources degraded and the states deviate from the desired manners. Determined by whether the open system has an energy exchange with the environment or not, the decoherence can be classified into the dissipation~\cite{doi10.114212402} and the pure dephasing~\cite{RevModPhys.73.357,PhysRevLett.100.037003}.
When a quantum system couples to an environment in a dephasing way, the quantum coherence of the system leaks to the environment, keeping its energy unchanged. In cold-atom systems, the random recoils of the atoms cause the dephasing process~\cite{Diehl2008,PhysRevA.94.023612}. By heating the atomic gas, the enhanced random collisions among atoms give rise to a severe dephasing process. The dissipation occurs in the spontaneous emission of a two-level atom~\cite{PhysRevA.68.043805,PhysRevA.62.013805}. In the electron-nuclear spin systems, the central electron spin generally experiences a dissipative process induced by the surrounding phonons and a flip-flop interaction with the nuclear spins~\cite{RevModPhys.85.79,doi:10.1126/science.1220513}. Compared with that of dephasing, the dissipative noise can lead to not only the leakage of quantum coherence, but also the relaxation of its energy.

A widely used approximation is the Born-Markovian approximation. It was found that the Born-Markovian dephasing noise forces the HL precision in Ramsey spectroscopy not only to return to the SNL at an optimal encoding time but also to become divergent in the long-time condition \cite{PhysRevLett.79.3865}. In the optical Mach-Zehnder interferometer-based quantum metrology scheme, photon dissipation also makes the metrology precision using entanglement~\cite{PhysRevLett.102.040403,PhysRevA.80.013825,PhysRevLett.107.083601} and squeezing~\cite{PhysRevA.81.033819,PhysRevA.95.053837} returns to or even worse than SNL. Having been proven to be universal for any Born-Markovian noises, these two destructive consequences are called the no-go theorem of noisy quantum metrology~\cite{PhysRevLett.116.120801,Albarelli2018restoringheisenberg}. The no-go theorem is the main obstacle to the practical realization of quantum metrology. Thus, the determination of whether this no-go theorem is ostensible or fundamental and whether can it be overcome is highly desirable from both theoretical and experimental perspectives.

In this section, we discuss two types of environmental noises, namely dephasing and dissipative noises, on the schemes of quantum metrology.

\subsection{Noisy Ramsey spectroscopy}\label{squeezedecoh}
Many previous works have studied the behavior of disentanglement under the influence of decoherence~\cite{PhysRevA.81.052330,PhysRevA.84.022329,PhysRevA.77.022320,PhysRevA.82.052330,PhysRevA.77.032342,PhysRevLett.105.050403}. In the Markovian case, it is showed that the entanglement of two qubits interacting independently with either quantum or classical noise becomes completely disentangled in a finite time~\cite{PhysRevLett.93.140404}. Such a disentanglement timescale has the same order as the usual spontaneous lifetime~\cite{PhysRevLett.93.140404}. However, in the non-Markovian case, Ref.~\cite{PhysRevLett.99.160502} reported that non-Markovian effects influence the entanglement dynamics and may give rise to a revival of entanglement even after the complete disentanglement has been presented for finite time periods. Moreover, Ref.~\cite{PhysRevLett.108.160402} demonstrated that non-Markovianity can be used to enhance the steady-state entanglement in a coherently coupled dimer system subject to dephasing noises. These results suggest that the non-Markovian memory effect may be used to improve the performance of noisy quantum metrology. The possible non-Markovianity-assisted metrological schemes have been discussed in Refs.~\cite{PhysRevA.88.035806,PhysRevA.103.L010601,PhysRevA.102.032607,PhysRevLett.127.060501,PhysRevA.102.022618,PhysRevLett.109.233601,PhysRevApplied.17.034073}.

We first review the effects of the local dephasing noises on the Ramsey spectroscopy. Influenced by the local dephasing noises, the encoding dynamics of the Ramsey spectroscopy is governed by the Hamiltonian
\begin{equation}
    \hat{H}=\sum_j\{\Delta \hat{\sigma}_j^\dag\hat{\sigma}_j+\sum_k[\omega_k\hat{b}^\dag_{j,k}\hat{b}_{j,k}+g_k\hat{\sigma}^z_j(\hat{b}^\dag_{jk}+\hat{b}_{jk})]\},
\end{equation}where $\hat{b}_{j,k}$ is the annihilation operator of the $k$th mode for the noise felt by the $j$th TLS and $g_k$ is their coupling strength. The coupling can be further characterized by the spectral density $J(\omega)=\sum_k|g_k|^2\delta(\omega-\omega_k)$. It generally takes the Ohmic-family form
$J(\omega)=\eta \omega(\omega/\omega_c)^{s-1}e^{-\omega/\omega_c}$, where $\eta$ is a dimensionless coupling constant, $\omega_c$ is a cutoff frequency, and $s$ is the so-called Ohmicity parameter. The environment is classified into the sub-Ohmic for $0 <s< 1$, the Ohmic
for $s = 1$, and the super-Ohmic for $s > 1$. The initial state of the total system is $\rho_\text{T}(0)=\rho(0)\otimes {e^{-\sum_{j,k}{\beta\omega_k}\hat{b}^\dag_{j,k}\hat{b}_{j,k}}/Z}$, with $Z$ being the partition function. After tracing out the noisy degrees of freedom, we obtain the exact non-Markovian master equation satisfied by the reduced density matrix of the probe as
\begin{equation}
    \dot{\rho}(t)=\sum_j\{-i{\Delta\over2}[\hat{\sigma}_j^z,\rho(t)]+{\gamma(t)\over2}[\hat{\sigma}_{j}^z\rho(t)\hat{\sigma}_{j}^z-\rho(t)]\},\label{demas}
\end{equation}
where $\gamma(t)=4\int_0^t d\tau\int_0^\infty d\omega J(\omega)\coth({\beta\omega\over2})\cos[\omega(t-\tau)]$. Solving Eq. \eqref{demas} under the initial condition $\rho(0)=|\text{GHZ}\rangle\langle \text{GHZ}|$, we obtain
\begin{eqnarray}
    \rho(t)&=&{1\over2}[|g\rangle\langle g|^{\otimes N}+|e\rangle\langle e|^{\otimes N}\nonumber\\
    &&+(e^{-N[i\Delta t+\Gamma(t)] }|e\rangle\langle g|^{\otimes N}+\text{h.c.})],
\end{eqnarray}where $\Gamma(t)=\int_0^t\gamma(\tau)d\tau$. Repeating the process of Ramsey spectroscopy in Sec. \ref{rmsspc}, we obtain the measurement results 1 with $P_1={1\over 2}[1-\cos(N\Delta t)e^{-N\Gamma(t)}]$ and 0 with $P_0={1\over 2}[1+\cos(N\Delta t)e^{-N\Gamma(t)}]$. Then the maximal CFI is calculated as
\begin{equation}
    F_{\omega_0}(t)=(Nt)^2e^{-2N\Gamma(t)}
\end{equation}when $\Delta t=k\pi/2$ with $k$ being odd numbers. Then the corresponding frequency sensitivity under the repetitive experiments within the time duration $T$ reads
\begin{equation}
    \delta\omega_0(t)=\Big[{e^{2N\Gamma(t)}\over TN^2t }\Big]^{1/2}.\label{nmdesen}
\end{equation}
The minimal $\delta\omega_0$ is reached at the encoding time satisfying ${d\over dt}{e^{2N\Gamma(t)}\over t}=0$.

In the special case that the probe-noise coupling is weak and the time scale of the noisy correlation function is much smaller than the one of the system, we can make the Markovian approximation by extending the $t$ in $\gamma(t)$ to infinity. After choosing the Ohmic spectral density, we have
\begin{equation}
    \lim_{t\rightarrow\infty}\gamma(t)={8\pi\eta/\beta}\equiv\bar{\gamma}.\label{mardep}
\end{equation}
Substituting Eq. \eqref{mardep} into Eq. \eqref{nmdesen}, we find that, although decreasing with time in the short-time regime, $\delta\omega_0(t)$ tends to be divergent in the long-time regime. It means that the encoding time as a resource to enhance the sensitivity is destroyed by the Markovian dephasing noises. The short-time optimal $\delta\omega_0$ is
\begin{equation}
 \min\delta\omega_0=\Big({2\bar{\gamma}e\over TN }\Big)^{1/2}
\end{equation}when $t={1\over 2N\bar{\gamma}}$.
It means that the Markovian dephasing noises force the HL in Ramsey spectroscopy not only to return to the SNL at an optimal encoding time but also to become divergent in the long-time condition \cite{PhysRevLett.79.3865}. Being universal for any Markovian noises \cite{Fujiwara_2008,Demkowicz-Dobrzanski2012,PhysRevLett.102.040403,PhysRevA.80.013825,PhysRevLett.92.230801,PhysRevLett.107.113603,Kacprowicz2010}, these two destructive consequences are called the no-go theorem of noisy quantum metrology \cite{Albarelli2018restoringheisenberg,PhysRevLett.116.120801}.

\begin{figure}[tbp]
	\centering
	\includegraphics[width=0.95\columnwidth]{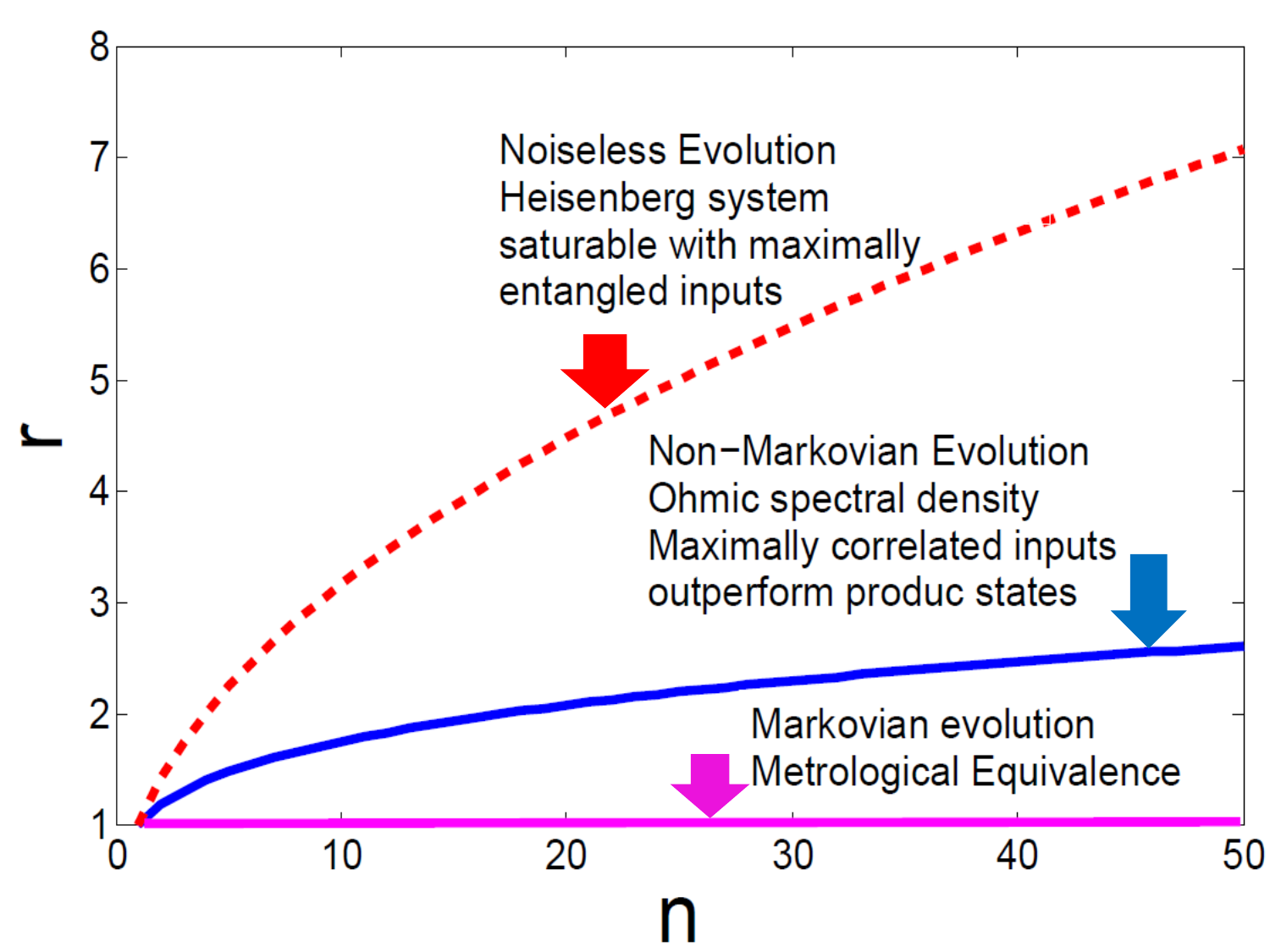}
	\caption{The ratio $r$ of the optimal sensitivities achieved with the product and the GHZ initial states is plotted as a function of the number of particles $N$. The dashed line shows the result in the absence of noise where $r=N^{1/2}$ (HL). The pink line is the Markovian result $r=1$. The blue line shows the non-Markovian result $r=N^{1/4}$ (Zeno limit) in the Ohmic spectral density case. Reproduced with permission \cite{PhysRevLett.109.233601}. Copyright 2012, American
Physical Society.}\label{W1}
\end{figure}

In the general non-Markovian case, the damping function $\Gamma(t)$ for the Ohmic spectral density at the zero temperature reads
\begin{equation}
    \Gamma(t)=4\eta \ln(1+\omega_c^2t^2).
\end{equation}Then, it is easy to derive from ${d\over dt}{e^{2N\Gamma(t)}\over t}=0$ that the minimal $\delta\omega_0$ is achieved at $t\simeq {1\over \omega_c\sqrt{8N\eta}}$ as \cite{PhysRevLett.109.233601}
\begin{equation}
    \min\delta\omega_0\simeq \Big({e\sqrt{8\eta}\over TN^{3/2}\omega_c}\Big)^{1/2}. \label{nonmadep}
\end{equation}
The scaling relation $\delta\omega_0\propto N^{-3/4}$ is called the Zeno limit. Equation \eqref{nonmadep} reveals that, being different from the Markovian case, the non-Markovian effect of the dephasing noises partially retrieves the quantum superiority of Ramsey spectroscopy, see Fig. \ref{W1}. However, the divergence fate of the precision in the long-time condition, i.e., $\lim_{t\rightarrow \infty}\delta\omega_0(t)=\infty$, does not change. A similar result on that the non-Markovian effect of the dephasing noises can reduce the HL to the Zeno limit $N^{-3/4}$ at an optimal time has been confirmed in Refs.~\cite{PhysRevA.84.012103,PhysRevLett.109.233601,PhysRevA.92.010102,PhysRevA.100.032318,PhysRevLett.116.120801}. In a recent experiment, this result was observed \cite{PhysRevLett.129.070502}.

Next, we consider the effects of the local dissipative noises on Ramsey spectroscopy~\cite{Wang_2017}. The encoding dynamics is governed by the Hamiltonian
\begin{equation}
\hat{H}=\sum_j\{\Delta\hat{\sigma}_{j}^\dag\hat{\sigma}_{j}+\sum_{k}[\omega_{k}\hat{b}_{j,k}^{\dagger}\hat{b}_{j,k}+g_{k}(\hat{b}_{j,k}\hat{\sigma}_{j}^{\dag}+\mathrm{h}.\mathrm{c}.)]\}.\label{eq:Hamiltonian}
\end{equation}
Assuming the environments are initially prepared in their vacuum states, the non-Markovian master equation is
\begin{eqnarray}
\dot{\rho}(t)&=&\sum_{j}\{{-i\omega(t)\over 2}[\hat{\sigma}_{j}^z,\rho(t)]\nonumber\\
&&+\frac{\gamma(t)}{2}[2\hat{\sigma}_{j}\rho(t)\hat{\sigma}_{j}^{\dag}-\{\rho(t),\hat{\sigma}_{j}^{\dag}\hat{\sigma}_{j}\}]\},
\end{eqnarray}
where $\gamma(t)+i\omega(t)=-\dot{u}(t)/u(t)$ with $u(t)$ determined by
\begin{equation}
u(t)+i\Delta u(t)+\int_{0}^{t}d\tau f(t-\tau)u(\tau)=0\label{eq:evolution-CF}
\end{equation}
under the initial condition $u(0)=1$. Here, $f(t-\tau)=\int_{0}^{\infty}d\omega J(\omega)e^{-i\omega (t-\tau)}$ is the environmental correlation function. If the initial state is a GHZ state, the reduced density matrix is given by
\begin{eqnarray}
\rho(t)&=&\frac{1}{2}\{(|g\rangle\langle g|)^{\otimes N}+[u(t)|e\rangle\langle g|]^{\otimes N}+[u(t)^{*}|g\rangle\langle e|]^{\otimes N}\nonumber\\
&&+[|u(t)|^{2}|e\rangle\langle e|+(1-|u(t)|^{2})|g\rangle\langle g|]^{\otimes N}\}.
\end{eqnarray}
Repeating the same process as the ideal case in Sec. \ref{rmsspc}, one can evaluate
\begin{equation}
\delta\omega_{0}(t)=\bigg{\{}\frac{T[\partial_{\omega_{0}}\text{Re}(u^{n}(t))]^{2}}{t[1-\text{Re}^{2}(u^{n}(t))]}\bigg{\}}^{-1/2}.\label{nmdsps}
\end{equation}

In the special case under the Markovian approximation, $u(t)\simeq e^{-\bar{\gamma}t-i(\Delta+\omega')t}$ with $\bar{\gamma}=\pi J(\omega_{0})$ and $\omega'=\mathcal{P}\int d\omega J(\omega)/(\omega-\Delta)$. Substituting it into Eq. \eqref{nmdsps} and optimizing $\Delta$ and $t$, one obtains
\begin{equation}
\min\delta\omega_{0}=\Big({e\bar{\gamma}\over NT}\Big)^{1/2}.
\end{equation}
Thus, being the same as the Markovian dephasing noises, the Markovian dissipation noises also force the metrology precision using the GHZ-type entanglement back to the SNL and thus the advantage of using entanglement in quantum metrology entirely disappears under the Markovian dissipative environments.

In the general non-Markovian case, a Laplace transform linearizes Eq.~\eqref{eq:evolution-CF} into $\tilde{u}(z)=[z+i\Delta+\int_0^\infty{J(\omega)\over z+i\omega}d\omega]^{-1}$. The solution of $u(t)$ is obtained by the inverse Laplace transform of $\tilde{u}(z)$, which can be done by finding its pole from~\cite{PhysRevA.81.052330,PhysRevResearch.1.023027}
\begin{equation}
y(E)\equiv\Delta-\int_0^\infty{J(\omega)\over\omega-E}d\omega =E,~(E=iz).\label{eigen}
\end{equation}
Note that the roots $E$ of Eq. \eqref{eigen} are just the eigenenergies of the total system formed by each atom and its local dissipative environment in the single-excitation space. Specifically, expanding the eigenstate as $|\Psi\rangle=(x\hat{\sigma}^{\dagger}+\sum_{k}y_{k}\hat{b}_{k}^{\dagger})|g,\{0_k\}\rangle$ and substituting it into $\hat{H}|\Psi\rangle=E|\Psi\rangle$ with $E$ being the eigenenergy, we have $(E-\omega_{0})x=\sum_{k}g_{k}y_{k}$ and $y_{k}=g_{k}x/(E-\omega_{k})$. They readily lead to Eq. \eqref{eigen}. Since $y(E)$ is a decreasing function in the regime $E < 0$, Eq. \eqref{eigen} has one and only one isolated root $E_b$ in this regime provided $y(0) < 0$. While $y(E)$ is ill-defined when $E>0$, Eq. \eqref{eigen} has infinite roots forming a continuous energy band in this regime. We call the eigenstate of the isolated eigenenergy $E_b$ bound state. After the inverse Laplace transform, we obtain
\begin{equation}
u(t)=Ze^{-iE_bt}+\int_{0}^{\infty}\frac{J(E)e^{-iEt}dE}{[E-\Delta-\delta(E)]^{2}+[\pi J(E)]^2},\label{eq:eq13}
\end{equation}
where $\delta(E)=\mathcal{P} \int_0^\infty \frac{J(\omega)}{E-\omega}d\omega $ with $\mathcal{P}$ being the Cauchy principal value. The first term with $Z=[1+\int_0^\infty{J(\omega)d\omega\over(E_b-\omega)^2}]^{-1}$ is from the bound state, and the second one is from the band energies. Oscillating with time in continuously changing frequencies, the integral tends to zero in the long-time condition due to the out-of-phase interference of the infinite components characterized by different oscillation frequency $E$. Thus, if the bound state is absent, then $\lim_{t\rightarrow\infty} u(t)= 0$ characterizes a complete dissipation, while if the bound state is formed, then $\lim_{t\rightarrow\infty} u(t)=Ze^{-iE_b t}$ implies a suppressed dissipation. It can be evaluated that the bound state is formed for the Ohmic-family spectral density when $\omega_0<\eta\omega_c\underline{\Gamma}(s)$, where $\underline{\Gamma}(s)$ is the Euler's gamma function.

By substituting the form $Ze^{-iE_bt}$ for the large-time $u(t)$ in the presence of the bound state into Eq. (\ref{nmdsps}), we have
\begin{equation}
\min(\delta \omega_0)= Z^{-(N+1)}(N^2Tt)^{-1/2},\label{bdsprc}
\end{equation}where the dependence of $E_b$ on $\omega_0$ has been considered via $\partial_{\omega_0}E_b=Z$. It is found that the bound state causes the encoding time as a metrology resource to be completely recovered. The precision asymptotically approaches the ideal HL (\ref{RIdeal}) for $N\ll \lfloor -1/\ln Z\rfloor$ when $Z$ reaches unity. Therefore, whether the superiority of the quantum metrology under dissipative noises exists or not highly depends on the formation of the bound state and the non-Markovian effect. When the decoherence is Markovian or the bound state is absent, the quantum superiority is destroyed; whereas when the decoherence is non-Markovian and the bound state is formed, the quantum superiority is retrieved. The result suggests a guideline for experimentation to implement the ultrasensitive measurement in the practical noise situation by engineering the formation of the bound state. This could be realized by the technique of quantum reservoir engineering \cite{doi:10.1126/science.1261033}. The bound state and its role in the non-Markovian dynamics have been observed in circuit QED \cite{Liu2016} and ultracold atom \cite{Kri2018,Kwon2022} systems.

\subsection{Noisy Mach-Zehnder interferomer}\label{entangledecoh}
Photon dissipation is the main decoherence source in the Mach-Zehnder interferomer. Previous work dealt with photon dissipation by phenomenologically introducing incomplete transmission coefficients in beam splitters~\cite{PhysRevLett.102.040403,PhysRevA.81.033819,PhysRevA.90.033846,PhysRevA.95.053837,PhysRevLett.108.130402}, which is equivalent to the Born-Markovian approximation description. Although the calculation under the Born-Markovian approximation is much more convenient, it may lose important physics. To reveal the actual performance of the scheme, the effect of photon dissipation on the scheme should be studied by discarding the widely adopted Born-Markovian approximation.

Taking the local dissipative noise of the encoding optical path into account, the encoding dynamics of the Mach-Zehnder interferometer is governed by
\begin{eqnarray}
\dot{\rho}(t)&=&-i[\omega_0\hat{a}_1^\dag\hat{a}_1+\Omega(t)\hat{a}^\dag_1\hat{a},\rho(t)]\nonumber\\
&&+\kappa(t)[2\hat{a}_2\rho(t)\hat{a}^\dag_2-\{\rho(t),\hat{a}_2^\dag\hat{a}_2\}],
\end{eqnarray}where $\kappa(t)-i\Omega(t)=-\dot{c}(t)/c(t)$. $c(t)$ is determined by
\begin{equation}\dot{c}(t)+i(\gamma+\omega_0)c(t)+\int_{0}^{t}f(t-\tau)c(\tau)d\tau=0\label{xtmax}
\end{equation}under $c(0)=1$.
Repeating the similar procedure as Sec. \ref{sbmzid} leads to  \cite{PhysRevLett.123.040402}
\begin{eqnarray}
	\bar{M}&=& \text{Re}[c(t)e^{i\omega_{0}t}](\sinh^{2} r -|\alpha|^2),\nonumber\\
	\delta M^2 &=& \text{Im}[c(t)e^{i\omega_{0}t}]^{2} [|\alpha \cosh r-\alpha^{\ast} \sinh r e^{i \phi}|^2 +\sinh^{2}r]\nonumber\\
&&+\text{Re}[c(t)e^{i\omega_{0}t}]^{2}[|\alpha|^{2}+\frac{1}{2} \sinh^{2}(2r)]+ \frac{1-|c(t)|^2}{2} N.\nonumber
\end{eqnarray}

Applying the Markovian solution $c_\text{MA}(t)=e^{-[\kappa+i(\omega_0+\gamma+\Delta)]t}$ with $\kappa=\pi J(\omega_0+\gamma)$ and $\Delta=\mathcal{P}\int_0^\infty{J(\omega)\over\omega_0+\gamma-\omega}d\omega$, we obtain $\min\delta \gamma\simeq({e^{2\kappa t}-1\over2 N t^2})^{1/2}$ when $\beta=(2\sqrt{N})^{-1}$ and $ \varphi=2\phi$. Getting divergent in the long-time limit, it's minimum at $t=\kappa^{-1}$ returns the SNL $e\kappa(2N)^{-1/2}$. Thus, the quantum superiority of the scheme in the Markovian noise disappears completely, which is consistent with the result in the Ramsey spectroscopy.

\begin{figure}[tbp]
	\centering
	\includegraphics[height=.44\columnwidth]{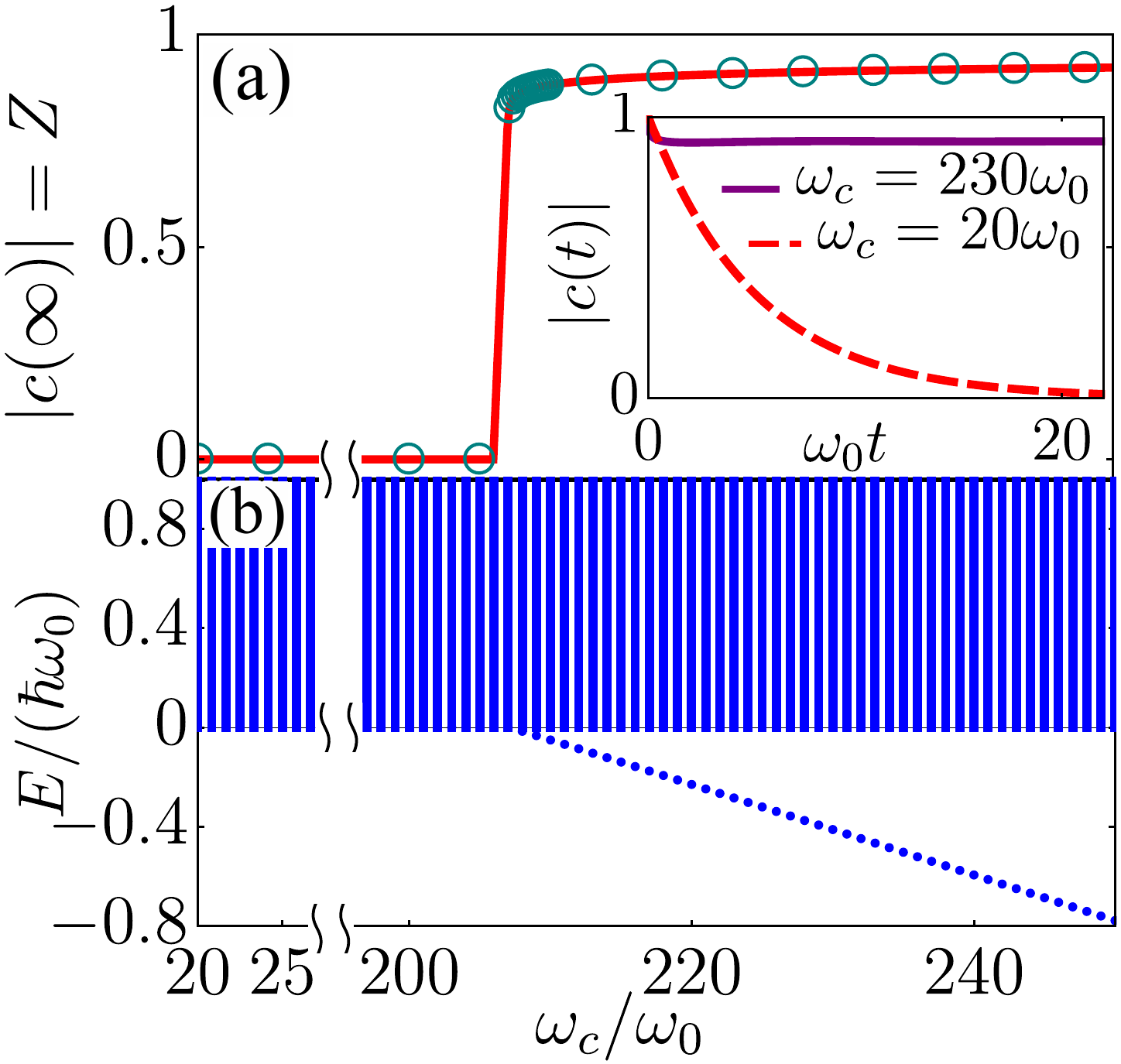}~\includegraphics[height=.44\columnwidth]{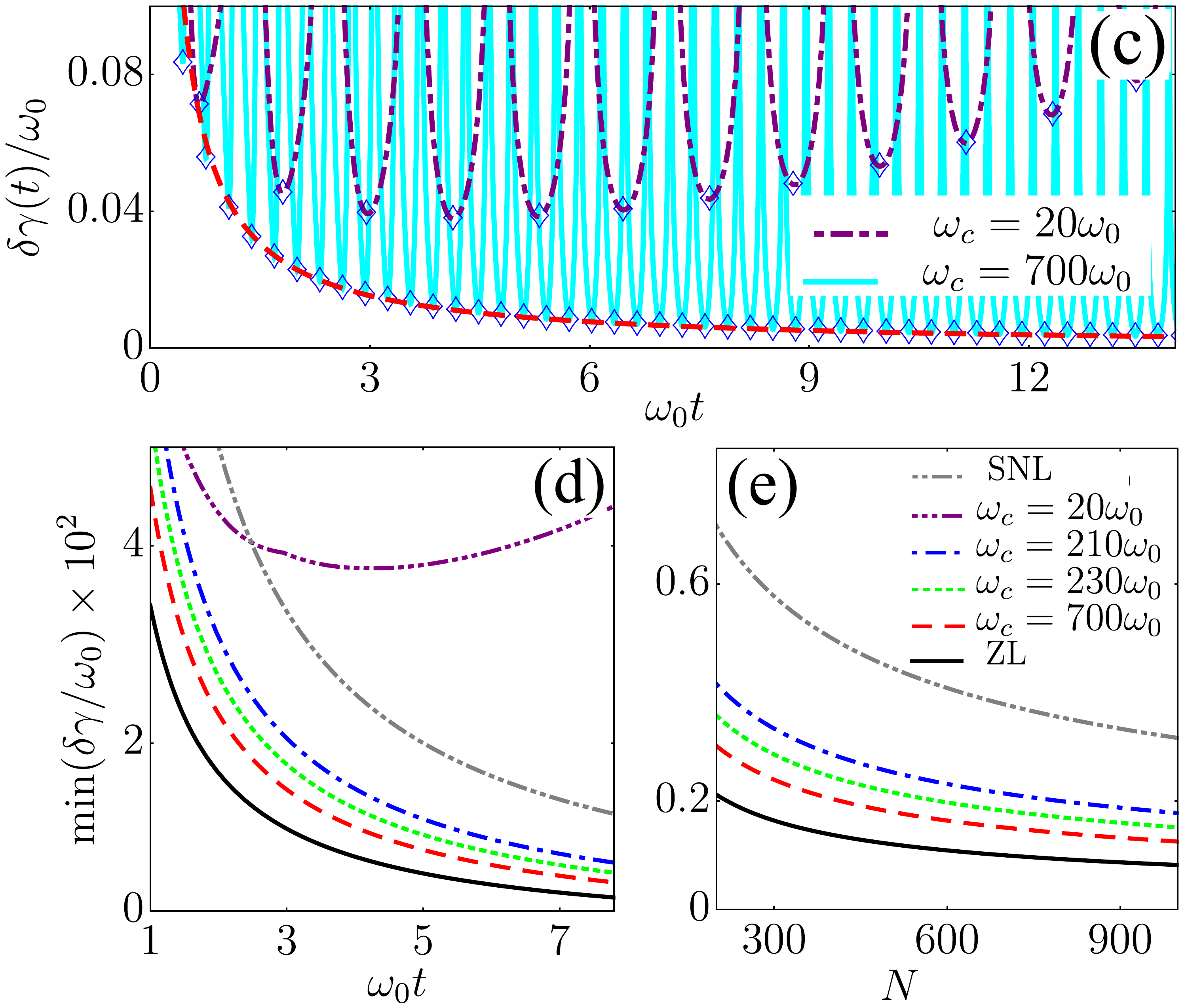}\\
	\caption {(a) $|c(\infty)|$ (dark cyan circles) by solving Eq. \eqref{xtmax}, which coincides with $Z$ (red solid line) from the bound state. The inset shows $|c(t)|$. (b) The energy spectrum of the whole system of the optical field and its environment. $\gamma=\pi\omega_0$ and $\eta=0.02$ are used. (c) Evolution of $\delta\gamma(t)$ without (purple dot-dashed line) and with (cyan solid line) the bound state, where the local minima match with the curve (red dashed line) of Eq. \eqref{J9}. Local minima of $\delta\gamma(t)$ as a function of time (d) and $N$ (e) in different $\omega_c$. $\iota=(2\sqrt{N})^{-1}$, $N=100$ in (d), and $t=10\omega_0^{-1}$ in (e) are used. Reproduced with permission \cite{PhysRevLett.123.040402}. Copyright 2019, American Physical Society. }
	\label{MZI}
\end{figure}

In the general non-Markovian dynamics, similar analysis as Sec. \ref{squeezedecoh} results in that, as long as $\omega_0+\gamma-\eta\omega_c\underline{\Gamma}(s)\leq0$, a bound state between the second optical field and its environment is formed and thus $\lim_{t\rightarrow \infty}c(t)=Ze^{-i\varpi_\text{b}t}$. Focusing on the case in the presence of the bound state and substituting the asymptotic solution $Ze^{-i\varpi_\text{b}t}$ into the result of $\delta\gamma$, one can obtain
\begin{equation}\label{J9}
	\min\delta\gamma|_{\iota=(2\sqrt{N})^{-1}}=\frac{(tN^{3/4})^{-1}}{Z}\bigg{(}1+\frac{1-Z^2}{2Z^2}N^{\frac{1}{2}}\bigg{)}^{\frac{1}{2}},
\end{equation}
when $t=\frac{(2m+1)\pi }{2|\omega_0-\varpi_\text{b}|}$ and $\varphi=2\phi$. Equation \eqref{J9} remarkably reveals that, even in the long-time condition, $\delta\gamma$ asymptotically tends to the ideal Zeno limit with $Z$ approaching 1, which is controllable by manipulating the spectral density $J(\omega)$ and the working frequency $\omega_0$ of the probe. It is in sharp contrast to the phenomenological and the Markovian approximate ones, which become divergent with increasing time. Figures \ref{MZI}(a) and \ref{MZI}(b) confirm that the formation of the bound state makes  $|c(\infty)|$ occur an abrupt change from zero to finite values exactly coinciding with $Z$. Figure \ref{MZI}(c) indicates that, as long as the bound state is formed, the profile of the local minima gets to be a decreasing function with time. Thus, the bound state makes the superiority of the encoding time as a resource in the ideal metrology case recovered. With the increase of $Z$ accompanying the increase of $\omega_c$, see Figs. \ref{MZI}(d) and \ref{MZI}(e), $\min(\delta\gamma)$ gets nearer and nearer the Zeno limit. All the results confirm that the precision asymptotically matching the analytical scaling \eqref{J9} approaches the Zeno limit with the formation of the bound state. Thus, the non-Markovian effect and the formation of a bound state between the quantum probe and its environment are two essential reasons for retrieving the Zeno limit: the bound state supplies the intrinsic ability and the non-Markovian effect supplies the dynamical way.

\subsection{Noisy quantum critical metrology }\label{criticdecoh}

Quantum criticality has been presented as a novel and efficient resource to improve the performances of quantum sensing and quantum metrology. Generally, protocols of criticality-based quantum metrology often work without decoherence. However, the many-body systems, which experience quantum phase transitions, usually interact with their surrounding environments. In this sense, the effect of decoherence should be taken into account. In Ref.~\cite{wantinghe}, the authors discussed the influence of photon relaxations on the inverted variance of quantum-Rabi-model-based quantum metrology around the quantum phase transition. They found that the achieved precision still diverges when approaching the criticality, but the power-law exponent is smaller than that for the noise-free case. Recently, it has been reported that the $p$-body Markovian dephasing dynamics of $N$-spin GHZ states evolving under a $k$-body Hamiltonian showed an $N^{-k+p/2}$ scaling of the estimation error~\cite{PhysRevLett.119.010403}.

In Ref.~\cite{PhysRevA.104.L020601}, the authors investigated the influences of dephasing on quantum critical metrology. In their scheme, a one-dimensional transverse-field Ising model
\begin{equation}
\hat{H}_{0}=-J\sum_{i=1}^{N-1}\hat{\sigma}_{i}^{z}\hat{\sigma}_{i+1}^{z}-B\sum_{i=1}^{N}\hat{\sigma}_{i}^{x}
\end{equation}
is prepared in the ground state at the quantum critical point. Then a small field $h$ is applied as $h\sum_{i=1}^{N}\hat{\sigma}_{i}^{z}$. After the free evolution of time $t$, a quantum measurement is performed and the result is compared with that obtained without applying the field $h$. The sensitivity is defined as the smallest $h$ that yields a measurement difference greater than the quantum fluctuation for an evolution time $t$, i.e., $\eta_{h}=h_{\min}\sqrt{t}$. The sensitivity has a theoretical lower bound $\eta_{h}\geq1/\sqrt{\mathcal{F}_ht}$, where the QFI $\mathcal{F}_h$ is related to the spectral function $\chi''(\omega)=\pi N\sum_{n\neq0}|\langle 0|\hat{M}|n\rangle|^{2}\delta(\omega-E_{n})$ by
\begin{equation}
\mathcal{F}_h=\frac{8N}{\pi}\int d\omega\chi''(\omega)\frac{1-\cos(\omega t)}{\omega^{2}}.
\end{equation}
Here $\hat{M}=\frac{1}{N}\sum_{i=1}^{N}\hat{\sigma}_{i}^{z}$. In the ideal noiseless case, one can find $\mathcal{F}_h\propto N^{15/4}$ in the long-time limit $t>\xi$ with $\xi$ being the correlation length. If $t<\xi$, the scaling relation becomes $\mathcal{F}_h\propto t^{2}N^{7/4}$. Taking the effect of decoherence into account, the authors couple each spin of the Ising chain to an independent bosonic environment. The environment temperature is set to be zero and the noise spectrum is chosen as the Ohmic spectral density. They found
\begin{equation}
\mathcal{F}_h\propto Nt^{2}.
\end{equation}
This result means that the local dephasing noises force the ideal super-HL scaling back to the SNL. Such a negative effect still holds for the non-Ohmic spectral densities.

\section{Decoherence Control}\label{control}
In this section, we give a brief review of the decoherence control schemes to minimize the unwanted effects of decoherence in metrological schemes.

\subsection{Dynamical control}

\begin{figure}[tbp]
	\centering
	\includegraphics[width=0.8\columnwidth]{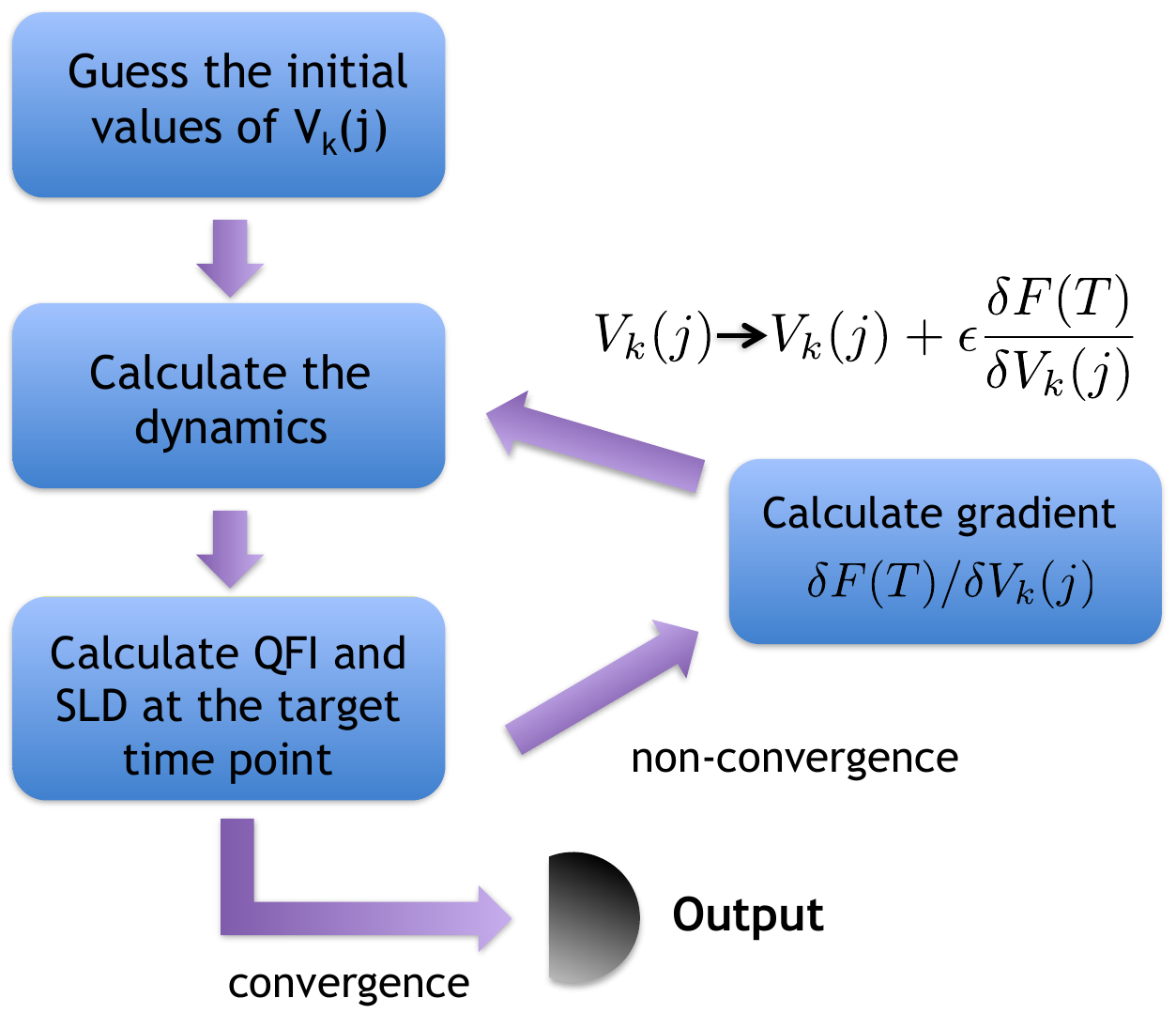}
	\caption{Flow chart of the optimal control algorithm.  Reproduced with
permission \cite{PhysRevA.96.012117}. Copyright 2017, American Physical Society. }\label{W2}
\end{figure}

Dynamical control is a mainstream to suppress the decoherence in quantum metrology. Under the Markovian approximation, Refs.~\cite{KHANEJA2005296,PhysRevA.103.042615,PhysRevA.96.012117,PhysRevA.102.012208,https://doi.org/10.1002/qute.202100080} proposed a gradient ascent pulse engineering scheme to beat the noisy effect on quantum metrology. It resorts to applying a series of control fields $\sum_{k}V_{k}(t)\hat{H}_{k}$ on the probe to overcome the decoherence effect. The amplitudes $V_{k}(t)$ of the control fields are optimized by the algorithm depicted in Fig. \ref{W2}. It has been demonstrated in the single-atom system that the precision limit under the controlled scheme can go beyond the constraints put by the coherent time. How to generalize this scheme to $N$-body quantum metrology is an open question. Going beyond the Markovian approximation, the dynamical decoupling method is a popular way to fight against the dephasing noise. Reference~\cite{PhysRevA.94.052322} used a dynamical decoupling method to beat the dephasing noises. It successfully revives the scaling of the measurement precision to $N^{-k}$ with $5/6\leq k\leq 11/12$. Such scaling is beyond the SNL and very close to the HL. A similar scheme was reported to beat the dephasing in quantum metrology using spin squeezing \cite{PhysRevA.89.063604}. These results convincingly demonstrated that the dynamical decoupling method is a powerful tool to regain a high-precision sensitivity in the dephasing environment~\cite{PhysRevLett.98.100504,PhysRevLett.102.120502,PhysRevLett.100.160505,PhysRevLett.101.180403,PhysRevLett.102.247601,PhysRevLett.109.070502,PhysRevLett.110.156402}. The dynamical decoupling method also was used to beat the dissipative noises. Reference~\cite{PhysRevA.87.032102} proposed a scheme to enhance the precision of parameter estimation in dissipative systems by employing dynamical decoupling pulses. It was found that the $N$-qubit dissipative systems can be preserved in the HL by applying a series of ideal $\pi$ pulses on the qubits. A Floquet engineering strategy was proposed to overcome the no-go theorem of noisy quantum metrology \cite{PhysRevLett.131.050801}. It is found that, by applying a periodic driving on the atoms of the Ramsey spectroscopy, the ultimate sensitivity to measure their frequency characterized by the QFI returns to the ideal $t^2$ scaling with the encoding time $t$ whenever a Floquet bound state is formed by the system consisting of each driven atom and its local dissipative noise. Combined with the optimal control, this mechanism also makes the ideal HL scaling with the atom number $N$ retrieved by optimizing the driving parameters. Simultaneously retrieving the quantum superiority of the encoding time and the number of entangled particles, the scheme supplies an efficient way to overcome the no-go theorem of noisy quantum metrology.

\subsection{Quantum error correction}
Quantum error correction plays an important role in the realization of high-precision quantum metrology in the presence of noises \cite{PhysRevLett.112.080801,PhysRevLett.112.150802,Lu2015,PhysRevLett.116.230502,Reiter2017,Zhou2018,PhysRevLett.128.140503}. To beat the local dephasing noises of Ramsey spectroscopy, Ref. \cite{PhysRevLett.112.080801} proposed a scheme of using $m$ physical qubits to encode a logical qubit $|0_L\rangle=(|+\rangle^{\otimes m} +|-\rangle^{\otimes m})/\sqrt{2}$ and $|1_L\rangle=(|+\rangle^{\otimes m} -|-\rangle^{\otimes m})/\sqrt{2}$, where $\hat{\sigma}_x|\pm\rangle=\pm|\pm\rangle$. The encoding dynamics under the influence of dephasing is governed by
\begin{equation}
\dot{\rho}(t)=-i \lambda [\hat{H},\rho(t)]+\sum_{j=1}^{m} \frac{\gamma}{2}[ \hat{\sigma}_j^z \rho(t) \hat{\sigma}_j^z  -\rho(t)], \label{ms}
\end{equation}
which causes a phase-flip error to any one of the physical qubits in a probability $1-p$, with $p=(1+e^{-\gamma t})/2$. Such kind of phase-flip error can be corrected as follows if the qubits occurring the error is fewer than $(m-1)/2$. First, we make a measurement, which projects $\rho(t)$ to the basis $\{\sigma_z^{{\bf k}}|+\rangle^{\otimes m},\sigma_z^{{\bf k}}|-\rangle^{\otimes m}\}$, with $\sigma_z^{\bf k}=\hat{\sigma}_z^{k_1}\otimes \cdots\otimes \hat{\sigma}_z^{k_m}$ and $k_j\in\{0,1\}$, to diagnose which qubits occur the error. Without changing $\rho(t)$, the measurement gives a result $1$ for $k_j=1$, while keeping other $k_{j'\neq j}$ zero, called error syndrome if the $j$th qubit occurs the error. Second, the error is completely corrected by applying an operation $\hat{\sigma}_z^{\bf k}$ obtained from the error syndrome ${\bf k}$. On the other hand, if the qubits occurring the phase-flip error are more than $(m-1)/2$, we judge that a phase-flip error occurs at the logic-qubit level and performs a correction operation on the logic qubit with probability $1-p_L$. Then the state is described $\mathcal{E}^{L}(p_L)\rho(t)=p_L\rho(t)+(1-p_L)\hat{\sigma}_z^{(L)}\rho(t)\hat{\sigma}_z^{(L)}$, where $\hat{\sigma}_z^{(L)}$ acting on the states of  the logic qubits and
\begin{equation}
    p_L=\sum_{k=0}^{(m-1)/2}\binom{m}{k} p^{m-k}(1-p)^k.
\end{equation}
The expansion of $p_L$ for small $1-p$ results in  $1-p_L\simeq \binom{m}{(m+1)/2}(1-p)^{(m+1)/2}$, which reveals an exponential suppression to the noise rate at the logic-qubit level.

Now, consider a metrology scheme using $Nm$ physical qubits. After the free evolution, the state becomes $|\psi_\lambda^L\rangle=(e^{-iN\theta_\lambda /2}|0_L\rangle ^{\otimes N}+e^{iN\theta_\lambda /2}|1_L\rangle ^{\otimes N})/\sqrt{2}$. Then, the state is subjected to local phase noises acting on all $Nm$ physical qubits. After performing the above error corrections to each block of $m$ qubits, the phase noise at the logic-qubit level is reduced and the state becomes $\rho_\lambda^L=[\mathcal{E}^L(p_L)]^{\otimes N}|\psi_\lambda^L\rangle\langle\psi_\lambda^L|$. It is readily calculated that the QFI of $\rho_\lambda^L$ is
\begin{equation}
    \mathcal{F}_\theta = (2p_L-1)^{2N} N^2.
\end{equation}
By using the approximation $\binom{m}{(m+1)/2}<2^m$, it can be shown that $(2p_L-1)^{2N}\rightarrow 1$ and $\mathcal{F}_\theta \approx N^2$ for $N\rightarrow\infty$ as long as $4N(2\sqrt{1-p})^m\ll1$, i.e., $m\sim O(\log N)$ \cite{PhysRevLett.112.080801}. Thus, the QFI of Ramsey spectroscopy under the influence of local noises is stabilized and the HL is attained, with only a logarithmic overhead.

Without resorting to measurements and operations, Ref. \cite{Reiter2017} developed an always-on dissipative quantum error correction to both the spin- and phase-flip errors and applied it in noisy quantum sensing in a trapped-ion system. To correct the spin-flip error, the logic qubit is encoded in three physical qubits, i.e., $|\psi\rangle=a|0_L\rangle+b|1_L\rangle=a|000\rangle+b|111\rangle$. The spin-flip error is caused by $\mathcal{L}_\text{noise}(\rho)=\sum_{k=1}^3\mathcal{D}[\hat{L}_k](\rho)$ with $\hat{L}_{k}=\sqrt{\Gamma}\hat{\sigma}_k^x$. Via interrogating the two-body stabilizer operators $\hat{S}_{ij}=\hat{\sigma}_i^z\hat{\sigma}_j^z$, one can find which qubit occurs the single-flip error. If the second physical qubit occurs a spin-flip error, then $
\hat{S}_{12}|\psi\rangle=\hat{S}_{23}|\psi\rangle=-|\psi\rangle$. The error in the first and the second qubits can be found in an analogous fashion. The recovery protocol is realized in a continuous manner by the dissipative dynamics
$\mathcal{L}_\text{qec}=\sum_{k=1}^3\mathcal{D}[\hat{L}_\text{qec}^{(k)}]$, where $\hat{L}_\text{qec}^{(2)}=\sqrt{\Gamma_\text{qec}}\hat{\sigma}^x_2{1-\hat{S}_{12}\over 2}{1-\hat{S}_{23}\over 2}$, $\hat{L}_\text{qec}^{(1)}=\sqrt{\Gamma_\text{qec}}\hat{\sigma}^x_1{1-\hat{S}_{21}\over 2}{1-\hat{S}_{13}\over 2}$, and $\hat{L}_\text{qec}^{(3)}=\sqrt{\Gamma_\text{qec}}\hat{\sigma}^x_3{1-\hat{S}_{13}\over 2}{1-\hat{S}_{32}\over 2}$. Then the total dynamics is governed by
\begin{eqnarray}
    \dot{\rho}(t)=-i[\hat{H},\rho(t)]+\mathcal{L}_\text{noise}(\rho(t))+\mathcal{L}_\text{qec}(\rho(t)).
\end{eqnarray}
By replacing $|0\rangle\rightarrow |+\rangle$, $|1\rangle\rightarrow|-\rangle$, and $\hat{\sigma}_z\leftrightarrow\hat{\sigma}_x$, the above scheme is applicable to correct the phase-flip error. Thus, both the phase- and spin-flip errors can be well corrected by this continuous-performing scheme without disturbance of the measurements and feedback operations.

Up to now, all the schemes on the quantum error correction are established under the Markovian approximation. How to develop the quantum error correction for non-Markovian noises is still an open question.

\subsection{Nondemolition measurement }
It has attracted much attention to avoid the destructive impacts of the noises on quantum metrology via carefully designing measurement schemes. Reference \cite{PhysRevLett.111.123601} proposed an asymmetric Ramsey technique to protect the atoms in Ramsey spectroscopy from spontaneous emissions. It complements the normal $\pi/2$ pulse with a phase distribution pulse to prepare the states protected from collective decoherence. Reference \cite{PhysRevLett.111.090801} proposed an adaptive scheme with nondemolition (weak) measurements to avoid the noisy effect on Ramsey spectroscopy using spin squeezing. A general adaptive quantum metrological scheme to beat the Markovian noises was presented in Ref. \cite{PhysRevX.7.041009}. The total evolution time is divided into a number of steps interleaved with general unitary controls. The collective measurement performed in the end allows us to regard the scheme as a general adaptive protocol where measurement results at some stage of the protocol influence the control actions applied at later steps.

References \cite{Albarelli2018restoringheisenberg,PhysRevLett.125.200505} proposed a continuous quantum nondemolition measurement of an atomic ensemble to eliminate the destructive impacts of independent dephasing acting on each atom. Consider that an ensemble of $N$ two-level atoms is
rotating around the $z$ axis with an angular frequency $\omega$. The aim is to estimate $\omega$. Each atom is  equally and independently subjected to a Markovian
dephasing. The atoms are prepared in a spin coherent initially. Continuous monitoring of the collective spin operator $\hat{J}_y$ is performed such that the conditional dynamics of the ensemble is
described by the stochastic master equation
\begin{eqnarray}
d\rho_c=-i\omega[\hat{J}_z,\rho_c]dt+{\kappa\over 2}\sum_{j=1}^N[\hat{\sigma}_j^z\rho_c\hat{\sigma}_j^z-\rho_c]dt+\Gamma [\hat{J_y}\rho_c\hat{J}_y\nonumber\\
-\{\hat{J}_y^2,\rho_c\}/2]dt+\sqrt{\eta\Gamma}[\{\hat{J}_y,\rho_c\}-2\text{Tr}(\rho_c\hat{J}_y)\rho_c] dw,~
\end{eqnarray}
conditioned by the measurement photocurrent $dy_t=2\sqrt{\eta\Gamma}\text{Tr}(\rho_c\hat{J}_y)dt+dw$.
Here, $\Gamma$ is the $\hat{J}_y$-measurement strength, $\eta$ is the measurement efficiency, and $(dw)^2=dt$ is a Wiener process. The QFI is contributed by the continuous photocurrent $dy_t$ and the final measurement on $\rho_c$, i.e.,
\begin{equation}
\mathcal{F}_\omega=F_{y_t}+\sum_\text{traj}p_\text{traj}\mathcal{F}_\omega(\rho_c^\text{(traj)}),
\end{equation}
where $F_{y_t}$ is the CFI of $y_t$ and $\mathcal{F}_\omega(\rho_c^\text{(traj)})$ is the QFI of $\rho_c^\text{traj}$ corresponding to different trajectories. It was found that, even without an initial entanglement and in the presence of the local dephasing noises, the maximal $\mathcal{F}_\omega$ at an optimal time can surpass the SNL. In this scheme, the monitoring-induced dynamics generates the resourceful state simultaneously with the frequency encoding. It reveals that continuous monitoring by quantum nondemolition measurements is a practical and relevant tool to obtain a quantum enhancement in spite of decoherence.

These schemes are only applicable to the Markovian noises. Being similar to the quantum error correction to suppress decoherence, how to develop general measurement schemes to control the non-Markovian noises of quantum metrology is still an open question. On the other hand, although partially recovering the quantum superiority in its scaling with $N$, the divergence fate of the precision in the long-time condition does not change. Therefore, how to retrieve the quantum superiority of both the encoding time and atom number as resources simultaneously is also an open question.

\section{Conclusions and outlook}\label{conclusion}

Quantum metrology is a rapidly developing discipline in the second revolution of quantum science and technology. Whether and when it practically outperforms its classical counterpart in absolute sensitivity depends sensitively on one's ability to prepare the quantum resources in a scalable way and to suppress decoherence induced by different noises in a controllable way. In the NISQ era, with advancements in quantum-state preparation and control and mitigating decoherence errors, quantum metrology has been gradually pushed to be on the brink of transitioning from the laboratory to practical use. Various quantum resources and decoherence control schemes in noisy quantum metrology have been widely investigated. These studies supplied possible ways to overcome the no-go theorem of noisy quantum metrology and paved the way to its realization in practical decoherence situations. However, there are still many open problems. First, how to design the optimal measurement scheme that saturates the value of QFI for different quantum resources and encoding protocols? Second, how to dig more quantum resources and quantum states of matter that may bring quantum superiority in metrology? Third, how to construct universal control methods to suppress the decoherence, especially for the non-Markovian one, of noisy quantum metrology? How to apply quantum metrology to generate revolutionary innovations in more advanced technologies? The efficient exploration of these problems from fundamental physics is hopefully to prompt the further development of quantum metrology in the near future. It also might supply basic physical ideas to reinforce the application of quantum metrology in innovating different kinds of advanced technologies.

\section*{Acknowledgments}

The work is supported by the National Natural Science Foundation of China (Grants No. 12275109, No. 12205128, No. 11834005, and No. 12247101).

\bibliography{reference}

\end{document}